\documentclass[%
 reprint,
 superscriptaddress,
 amsmath,amssymb,
 aps,
 prb,
 longbibliography
]{revtex4-2}
\usepackage{graphicx}% Include figure files
\usepackage{dcolumn}% Align table columns on decimal point
\usepackage{bm}% bold math
\usepackage{mathtools}
\usepackage{float}
\usepackage[usenames,dvipsnames]{color}
\usepackage[colorlinks=true, citecolor=blue, linkcolor=blue, urlcolor=blue]{hyperref}
\usepackage{xr-hyper}
\usepackage{cleveref} %to use \cref command
\usepackage[nolist]{acronym}
\usepackage[utf8]{inputenc}
\usepackage{physics}

\usepackage[table]{xcolor} % For row coloring
\usepackage{booktabs}      % For professional-looking tables
\usepackage{multirow}      % For multirow cells
\usepackage{array}         % For better column formatting
\usepackage{tabularray}
\usepackage{nicematrix}

\newcommand\revise[1]{\textcolor{black}{#1}}

\newcommand{\affiliationMain}{%
 Institute for Theoretical Physics and Bremen Center for Computational Materials Science, University of Bremen, 28259 Bremen, Germany}

\newcommand{\affiliationSecond}{%
Max Planck Institute for the Structure and Dynamics of Matter,
Center for Free-Electron Laser Science (CFEL),
Luruper Chaussee 149, 22761 Hamburg, Germany}
\newcommand{\affiliationThird}{%
Fritz Haber Institute of the Max Planck Society, 14195 Berlin, Germany}
\newcommand{\affiliationFourth}{%
PSI Center for Scientific Computing, Theory and Data, Paul Scherrer Institute, 5232 Villigen PSI, Switzerland}
\newcommand{\affiliationFifth}{%
Department of Physics, University of Fribourg, 1700 Fribourg, Switzerland}
\newcommand{\affiliationSixth}{%
Initiative for Computational Catalysis (ICC) and Center for Computational Quantum Physics (CCQ) The Flatiron Institute, New York, NY 10010, USA}

\begin{document}

\title{Probing topological Floquet states in graphene with ultrafast terahertz scanning tunneling microscopy}

\author{Nils Jacobsen}
\email{jacobsen@uni-bremen.de}
\affiliation{\affiliationMain}
\affiliation{\affiliationSecond}

\author{Michael Schüler}
\affiliation{\affiliationFourth}
\affiliation{\affiliationFifth}

\author{Angel Rubio}
\affiliation{\affiliationSecond}
\affiliation{\affiliationSixth}

\author{Martin Wolf}
\affiliation{\affiliationThird}

\author{Melanie Müller}
\email{m.mueller@fhi-berlin.mpg.de}
\affiliation{\affiliationThird}

\author{Michael A. Sentef}
\email{sentef@uni-bremen.de}
\affiliation{\affiliationMain}
\affiliation{\affiliationSecond}

% Date (optional)
\date{\today}

\begin{abstract}
Floquet control of band topology is a central theme in ultrafast quantum materials science. Established experimental probes of light-induced topological states include ultrafast transport and time- and angle-resolved photoemission spectroscopy, each with important strengths but also well-known limitations. Here we propose ultrafast terahertz scanning tunneling microscopy (THz-STM) as a real space energy-resolved probe of Floquet physics. We show that THz-STM enables direct local detection of bulk Floquet gaps and distinct Floquet edge state signatures. We derive a nonequilibrium Green's-function formalism for time-dependent tunneling that directly extends standard STM theory and provides an intuitive interpretation of rectified ultrafast tunneling currents. We apply the approach to bulk graphene and graphene nanoribbons of variable width. For the bulk, we show that THz-STM provides direct spectroscopic access to Floquet-induced gap openings, and we contrast pulsed pump--probe protocols with the continuous-wave Floquet steady-state limit. For finite ribbons, we demonstrate time- and space-resolved imaging of Floquet-induced topological edge states and identify the ribbon-width scale below which edge state protection breaks down. We further show how band structures of graphene nanoribbons and Floquet chiral edge modes can be reconstructed via Floquet quasiparticle interference. Finally we demonstrate that time-reversal symmetry-breaking chiral impurities induce characteristic spatial THz-STM signatures and thus allow to directly probe Floquet edge-state chirality.
\end{abstract}

% Keywords (optional)

\maketitle

% Main text
\section{Introduction}
Advances in ultrafast light sources have enabled controlled access to nonequilibrium states of quantum materials \cite{basov_towards_2017, de_la_torre_colloquium_2021}.
Groundbreaking discoveries in the field range from light-induced magnetism \cite{radu_transient_2011, disa_polarizing_2020, wang_light-induced_2022} and ferroelectricity \cite{nova_metastable_2019, li_terahertz_2019} via charge-density wave order and superconductivity \cite{kogar_light-induced_2020, wandel_enhanced_2022}, \cite{fausti_light-induced_2011, mitrano_possible_2016, buzzi_photomolecular_2020,rowe_resonant_2023} to topological states of matter \cite{wang_observation_2013,mciver_light-induced_2020}.

Specifically, the paradigm of the Floquet topological insulator \cite{lopez-rodriguez_analytic_2008, oka_photovoltaic_2009,lindner_floquet_2011,kitagawa_transport_2011,rudner_anomalous_2013,perez-piskunow_floquet_2014,usaj_irradiated_2014,perez-piskunow_hierarchy_2015,oka_floquet_2019,rudner_band_2020} has spurred much experimental interest. 
The key idea is that periodically driven Dirac fermions can acquire a dynamical mass term that opens a gap and drives a topological phase transition.
First experimentally implemented in photonic \cite{rechtsman_photonic_2013} and cold-atom \cite{jotzu_experimental_2014} platforms, the experimental verification in quantum materials is much more challenging due to issues of heating, dissipation, and the lack of straightforward probing techniques \cite{aeschlimann_survival_2021}. 
%Sub-picosecond ultrafast transport was finally used to provide evidence for the light-induced anomalous Hall effect  \cite{mciver_light-induced_2020}. 
%However, the theoretical interpretation of the data turned out to be challenging \cite{sato_microscopic_2019, nuske_floquet_2020}. In related work, photo-induced currents were interpreted in a Floquet topological insulator scenario, but again theory was required to interpret the data in this scenario \cite{lesko_optical_2025}. Other studies focus on the continuous-wave/steady-state regime and propose transport measurements \cite{esin_quantized_2018} or employ tunneling probes of Floquet-Andreev states in graphene Josephson junctions \cite{park_steady_2022}.
Sub-picosecond ultrafast transport has provided evidence for a light-induced anomalous Hall response \cite{mciver_light-induced_2020}. 
However, a microscopic interpretation of the measured signal is subtle and has triggered extensive theoretical discussion \cite{sato_microscopic_2019, nuske_floquet_2020}. 
\revise{In related work, photoinduced currents were interpreted within a Floquet-topological-insulator scenario, with theory playing an essential role in relating the measured current response to the light-induced Berry curvature \cite{lesko_optical_2025}.} 
Complementary efforts target the continuous-wave, steady-state regime, including transport measurements\cite{liu_signatures_2025} and tunneling studies of Floquet--Andreev states in graphene Josephson junctions \cite{park_steady_2022}.

In time- and angle-resolved photoemission spectroscopy \cite{boschini_time-resolved_2024}, ideally suited to map photon-dressed band structures in bulk, the concomitant occurrence of final-state photodressing of photoemitted electrons, known as Volkov states \cite{saathoff_laser-assisted_2008}, complicates a simple interpretation of spectral features spaced by multiples of the photon frequency as Floquet sidebands. This is a key reason why a unique identification of Floquet states in graphene required careful polarization-dependent measurements that disentangle Floquet and Volkov contributions via their interference \cite{merboldt_observation_2025, choi_observation_2025}, and have also motivated alternative purely optical approaches to detecting Floquet states \cite{tiwari_first-principle-based_2025, tiwari_robust_2025}.  
Despite much progress, a real-space, energy-resolved probe that can directly detect Floquet gap openings and edge-mode signatures is still missing. 

Floquet topology manifests itself directly in real space through Floquet-induced chiral edge modes \cite{rudner_anomalous_2013,perez-piskunow_floquet_2014,perez-piskunow_hierarchy_2015,farrell_edge-state_2016,chen_observing_2020}. This naturally implies the need for a time-resolved real-space probe of Floquet states in quantum materials. Moreover, a further—and still often underappreciated—obstacle for Floquet engineering in solids is precisely the intrinsic inhomogeneity of real materials. Spatial variations due to carrier density \cite{chen_charged-impurity_2008,zhang_origin_2009,deshpande_spatially_2009}, local strain \cite{teague_evidence_2009, choi_effects_2010}, the presence of Moiré potentials or sublattice symmetry breaking \cite{brihuega_quasiparticle_2008,wang_gaps_2016,zhou_substrate-induced_2007} can alter the Dirac electronic structure in graphene and thereby affect the spectral signatures and topological properties induced by a Floquet drive.
As a result, even under nominally uniform illumination, the topological properties and spectroscopic signatures of Floquet-driven graphene might vary at the nanoscale. In particular, Floquet gaps may compete with trivial gaps arising from substrate- or defect-induced sublattice symmetry breaking, so that the formation of topological edge modes can become sensitive to the local microscopic environment. Consequently, Floquet topological regions may coexist with topologically trivial patches if the effective mass term changes sign locally. 

The aforementioned photoemission- and transport-based measurements typically average over micrometer-scale areas and therefore cannot resolve how Floquet band dressing persists \cite{bielinski_floquetbloch_2025} —or breaks down— in the presence of inhomogeneity on microscopic length scales. Scanning tunneling microscopy (STM) provides atomic-scale real-space imaging and, through its spectroscopic signal, essentially probes the sample's local density of states (LDOS) \cite{berthod_spectroscopic_2018}, i.e.\ the energy-resolved density of available electronic states at a given position. It combines atomic-scale real-space resolution with high spectroscopic energy resolution, providing a direct way to access the spatial inhomogeneity of Floquet topology and to characterize the associated edge modes at atomically sharp boundaries and defect bound states \cite{lovey_floquet_2016}. This motivates the development of theoretical and experimental real-space approaches to study Floquet-engineered quantum materials at the atomic-scale. 

\begin{figure*}
    \centering
    \includegraphics[width=\linewidth]{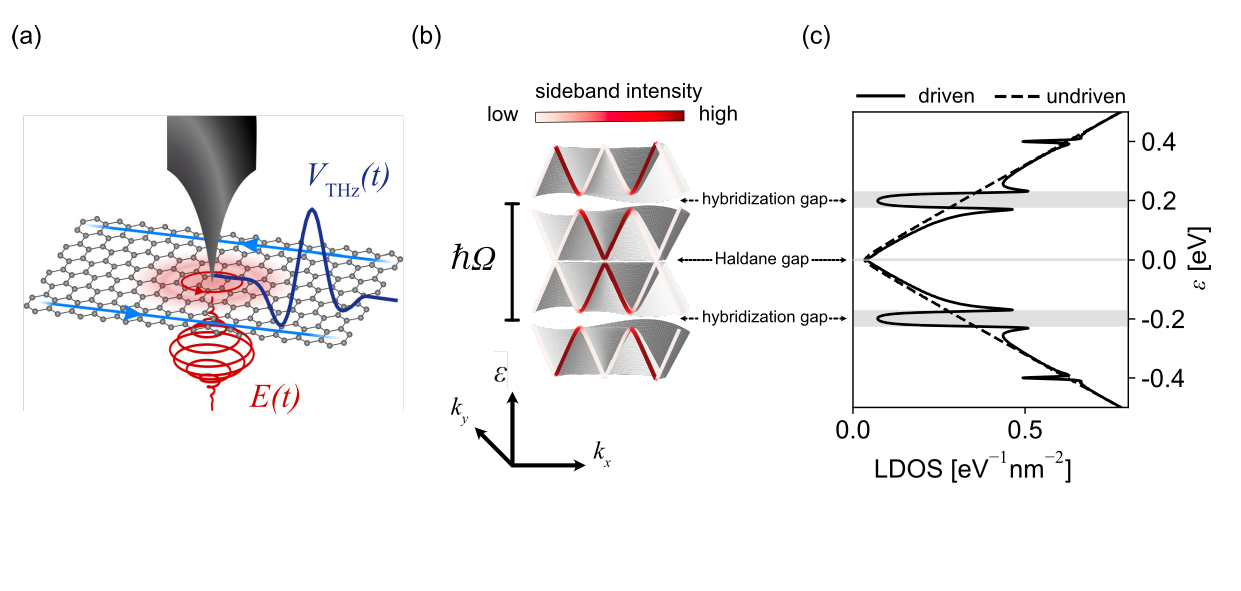}
    \vspace{-2cm}
    \caption{(a) Experimental geometry for measuring topological Floquet states in graphene by THz-STM. A circular pump pulse $E(t)$ excites a graphene nanoribbon under normal incidence. The in-plane drive creates a topological Floquet state with chiral edge modes on opposite edges that have opposite propagation directions. The state is probed using an ultrafast THz bias voltage $V_\mathrm{THz}(t)$ coupled to the STM tip. 
  (b) Floquet band structure around the $K$ point and 
  (c) corresponding LDOS in the bulk under circularly polarized driving. 
  The electric field strength is $E = 350~\mathrm{kV\,cm^{-1}}$ and the photon energy is $\hbar\Omega = 0.4~\mathrm{eV}$. 
  The band structure comprises Floquet replica bands shifted by integer multiples of the photon energy, with the spectral intensity shown along $k_x$ in the 3D plot. 
  Hybridization gaps appear at the edges of the Floquet Brillouin zone at $\varepsilon = \hbar\Omega/2$ and a Haldane gap opens at the Dirac point.}
    \label{fig:1-Floquet_bands+DoS}
\end{figure*}

Terahertz scanning tunneling microscopy (THz-STM) \cite{cocker_nanoscale_2021,muller_imaging_2023} combines real-space imaging with spectroscopic sensitivity to light-induced gap openings, thereby complementing momentum-space probes. Since its first implementation \cite{cocker_ultrafast_2013} and the demonstration of sub-cycle ultrafast control of tunneling \cite{cocker_tracking_2016, yoshioka_real-space_2016}, THz-STM milestones in probing ultrafast dynamics and nonequilibrium phases include tracking and controlling single molecule dynamics \cite{cocker_tracking_2016, peller_sub-cycle_2020, wang_atomic-scale_2022}, probing coherent phonon \cite{sheng_launching_2022,roelcke_ultrafast_2024,rai_influence_2025,lopez_atomic-scale_2025} and charge density wave \cite{sheng_terahertz_2024,lopez_atomic-scale_2025} dynamics locally at the angstrom scale, the real-space visualization of local potential variations and carrier dynamics \cite{jelic_ultrafast_2017,yoshida_subcycle_2019,yoshida_terahertz_2021}, THz-control of exciton formation in a single molecule \cite{kimura_ultrafast_2025}, or probing
THz-induced structural topological phase transitions \cite{jelic_terahertz_2025}. Furthermore, implementing THz scanning tunneling spectroscopy (THz-STS) approaches \cite{ammerman_lightwave-driven_2021,roelcke_ultrafast_2024} will expand the potential of THz-STM to study the ultrafast dynamics of the local density of states in quantum materials.

Here we propose THz-STM as a new tool to be added to the experimental toolbox for probing Floquet-engineered two-dimensional quantum materials and light-induced topological phase transitions. In order to showcase the opportunities that THz-STM opens up for imaging Floquet physics at the atomic scale, we first extend the general formalism to describe tunneling in STM to ultrafast tunneling currents by employing nonequilibrium Green's functions.
We derive a Floquet version of the well-known Meir-Wingreen formula \cite{wingreen_time-dependent_1993} that allows for an intuitive understanding of information contained in the tunneling current through a periodically driven material.
After a recapitulation of the salient features of Floquet topology in graphene, we apply our THz-STM formalism to bulk and ribbon graphene driven by continuous-wave and pulsed lasers.
We show that band gaps occurring under driving with circularly polarized light pulses in bulk graphene can be resolved by employing THz-STM as an ultrafast spectroscopic probe.
We further demonstrate the real-space imaging capabilities of THz-STM to probe topological edge states by studying finite-sized graphene nanoribbons of varying width, and show how the edge state protection eventually breaks down due to hybridization when the ribbons become too narrow. 
 Finally, we propose two distinct scattering probes of Floquet edge modes: (i) Floquet quasiparticle interference (Floquet QPI) as a tool to reconstruct edge state dispersions, and (ii) the circularly-dichroic Floquet LDOS renormalization around chiral impurities to identify edge mode chirality.

This paper is structured as follows. In the methods Sec.~\ref{sec:methods} we introduce the ultrafast THz-STM measurement protocol (Sec.~\ref{sec:thzstm}) and derive the nonequilibrium tunneling formalism (Sec.~\ref{sec:tunneling}), including its Floquet steady-state limit. In Sec.~\ref{sec:Floq_form} we briefly summarize Floquet engineering of band topology in graphene as relevant to what follows. In Sec.~\ref{sec:results} we present our results: we first analyze Floquet gap openings in bulk graphene (Sec.~\ref{sec:bulkgaps}) and compare continuous-wave and ultrafast pump--probe measurement protocols. We then demonstrate real-space spectroscopy of Floquet edge states in driven graphene nanoribbons (Sec.~\ref{sec:ribbons}). The results section concludes with Floquet QPI (Sec.~\ref{sec:qpi}) and the circularly-dichroic Floquet LDOS renormalization around chiral impurities (Sec.~\ref{sec:dichroism}). We end with a discussion of prospects arising from our predictions in Sec.~\ref{sec:discussion}. Technical details of the real-time implementation and the Floquet impurity T-matrix approach are provided in Appendices~\ref{appx:A} and \ref{appx:B}, respectively.

\section{Models and Methods}
\label{sec:methods}

\subsection{Ultrafast THz-STM}
\label{sec:thzstm}
We begin by briefly recalling conventional STM as a widely established tool for spectroscopic measurements of surfaces or two-dimensional materials \cite{berthod_spectroscopic_2018, avraham_quasiparticle_2018,yin_probing_2021}.
In standard STM, a static bias voltage $V_\mathrm{DC}$ is applied between the sample and a metallic tip and the tip is scanned across the sample.
At close distances, finite spatial overlap between tip and sample wavefunctions enables tunneling and gives rise to a current of the form 
\begin{align}
    I = -\frac{\mathrm{e}\Gamma}{\hbar} \int \mathcal{A}_{\vec{r}} (\varepsilon) \left[f(\varepsilon - \mu - \mathrm{e}V_\mathrm{DC} ) - f(\varepsilon -\mu) \right] \mathrm{d} \varepsilon,
\end{align}
% note on the sign convention: V > 0 means mu_sample = mu_tip - eV < mu_tip ie more occupied states in tip and tunneling from tip to sample as in standard convention for STM
where $\mathcal{A}_{\vec{r}}(\varepsilon)$ is the LDOS of the sample at energy $\varepsilon$ and the tip position $\vec{r}$, $f$ is the Fermi Dirac distribution, and $\mu$ is the chemical potential. %, $\mathrm{e}$ is the absolute value of the elementary charge and $\hbar$ the reduced Planck constant.
$\Gamma$ is a tunneling rate stemming from the tip density of states and tunneling matrix elements, whose energy dependence is typically omitted.
At low temperatures, due to the relation $\mathrm{d}f(\varepsilon)/\mathrm{d}\varepsilon \approx -\delta(\varepsilon)$, the tunneling conductance relates to the LDOS as
\begin{align}
   \frac {\mathrm{d}I}{\mathrm{d}V_\mathrm{DC}} \sim \mathcal{A}_{\vec{r}}( \mu + \mathrm{e}V_\mathrm{DC}) \label{eq:dIdV}.
\end{align}
Hence, STM can be used to probe the LDOS of the sample by sweeping the STM bias. 
This allows real-space spectroscopy with an energy resolution of millielectronvolts and a spatial resolution on the Ångstrom scale.
In the vicinity of defects, the LDOS is modulated due to interference between scattered electronic wave functions.
These so-called quasiparticle interference patterns allow to reconstruct energy dispersions and can, in principle, provide access to the possible topological structure of the electronic wave function through characteristic scattering signatures \cite{mallet_role_2012, avraham_quasiparticle_2018, yin_probing_2021}.

Ultrafast THz-STM extends the capabilities of conventional STM into the femtosecond time domain \cite{cocker_ultrafast_2013,cocker_nanoscale_2021,muller_imaging_2023}.
In THz-STM, the STM tip is illuminated with a single-cycle laser pulse, typically a half-cycle pulse in the THz frequency range.
The tip-enhanced THz electric field generates a time-dependent bias voltage $V_\mathrm{THz}(t)$ in the tunneling junction, which is added to the static DC bias. 
Whereas the temporal evolution $I(t)$ of the ultrafast tunneling current itself cannot be resolved due to the limited bandwidth of the STM electronics, the rectified charge
\begin{align}
   Q_\mathrm{rect} =\int_{-\infty}^\infty  I(t) \mathrm{d}t \label{eq:q_rect}
\end{align} 
is measured in experiments.
The time-resolution of THz-STM is thus set by the half-cycle duration of the probe pulse in the sub-picosecond regime.
$Q_\mathrm{rect}$ can be non-zero even if the temporal integral of the time-dependent bias vanishes owing to a non-linear and temporarily non-local current-bias dependence of the tunneling junction. 
The non-linearity is determined by $\mathcal{A}_{\vec{r}}$ in Eq.~\ref{eq:dIdV} and measured by scanning tunneling spectroscopy (STS).
Recent attempts are made to extend STS to ultrafast time scales \cite{ammerman_algorithm_2022, roelcke_ultrafast_2024}.

The temporal non-locality implies a memory effect of the tunneling currents of biases at previous times following from their nonequilibrium nature. 
This effect is typically ignored in the adiabatic interpretation of THz-STM experiments, where a stationary LDOS and a corresponding instantaneous tunneling conductance are assumed \cite{ammerman_algorithm_2022}. 
In a rigorous nonequilibrium description, as discussed in the following section, the notion of an LDOS is not generally well-defined out of equilibrium, so it is not a priori clear how static STM spectroscopy generalizes.

This open question is addressed in the following for the case of THz-STM measurements of topological Floquet states in graphene.
The geometric arrangement of the modeled experimental set-up is shown in Fig.~\ref{fig:1-Floquet_bands+DoS} a). 
The Floquet state is excited by a multicycle pump pulse $E(t)$ in the mid-infrared to optical frequency regime, which is circularly polarized in the graphene plane.
The pump pulse is assumed to excite the sample under normal incidence, so that only in-plane components of the Floquet driving field need to be considered. 
Specifically, for the present study, we neglect near-field effects and field enhancement of the pump pulse by the substrate and the STM tip, and instead assume the electronic states in the sample to be dressed by a plane-wave optical field.
The pump field is set to zero inside the vacuum tunneling gap.
Although such idealized conditions and normal-incidence excitation have not yet been experimentally realized, this setting provides a minimal theoretical model for our study. 
In particular, it allows us to neglect competing effects such as photon-assisted transport and optical lightwave-driven tunneling driven by out-of-plane components of the optical pump pulses.
The time-dependent THz field of the probe pulse is assumed to be polarized out-of-plane and to generate a transient bias voltage $V_\mathrm{THz}(t)$.

\subsection{Nonequilibrium tunneling formula} 
\label{sec:tunneling}
\subsubsection{Real-time formalism} 
In the following section, we sketch the derivation of the nonequilibrium tunneling formula. 
The result is an adaption of the Meir-Wingreen formula \cite{wingreen_time-dependent_1993, gianluca_stefanucci_robert_van_leeuwen_nonequilibrium_2013}, and has been applied to similar tunneling problems in previous studies \cite{liu_keldysh_2017,kwok_stm_2019,tuovinen_time-linear_2023}.

We model the ultrafast STM measurements within the nonequilibrium Green's function formalism.
Here, the propagation of charge carriers is described by the non-interacting Green's function $\hat G(t,t')$ with time arguments $t$ and $t'$ on the Keldysh contour. 
The Green's function evolves under the equation of motion
\begin{align}
\begin{split}
    \mathrm{i}\hbar \frac{\partial}{\partial t} &\hat G(t,t') - \hat h(t) \hat G(t,t') \\
    &= \int \mathrm{d}s \hat \Sigma(t,s)\hat{G}(s, t')  + \delta(t-t') \label{eq:eom1}
\end{split}
\end{align}
and its adjoint equation
\begin{align}
\begin{split}
    -\mathrm{i}\hbar \frac{\partial}{\partial t'} &\hat G(t,t') -  \hat G(t,t')\hat h(t') \\
    &= \int \mathrm{d}s \hat{G}(t, s)\hat \Sigma(s,t')  + \delta(t-t'). \label{eq:eom2}
\end{split}
\end{align}
Here, $\hat h(t)$ is the single particle Hamiltonian, which in this work is assumed as a tight-binding Hamiltonian, where a time-dependent electric pump field is included by a Peierls substitution, as described in the following section. 
$\hat \Sigma(t,t') =  \big(\hat\Sigma_\mathrm{t}+ \hat\Sigma_\mathrm{s}\big)(t,t')$ is the embedding self-energy taking into account the coupling to external reservoirs.
It decomposes into contributions $\hat\Sigma_\mathrm{t/s}(t,t')$ from the tip/substrate, which act as source and drain for the tunneling currents.
In a general setting, $\hat\Sigma(t,t')$ may also account for electronic correlations, which we omit in this work. 

The tunneling currents can be derived from the time evolution of the lesser component of the Green's function $ \hat G^<(t,t')$.
It relates to the density matrix as $\hat \rho(t) = -\mathrm{i}\hat G^<(t, t) $ and encodes electronic occupations of single-particle states in the sample.
Its derivative at equal times evaluates to
\begin{align}
\begin{split}
    \mathrm{i}\hbar \frac{\partial}{\partial t} \hat G^<(t, t) &= [\hat h(t),\hat G^<(t, t)] \\
    &+ \sum_{\alpha = \mathrm{t,s}}\bigg(\hat J_\mathrm{emb,\alpha}(t)  + \hat J^\dagger_\mathrm{emb,\alpha}(t)\bigg), \label{eq:dG<dt}
\end{split}
\end{align}
where the embedding terms $\hat J_\mathrm{emb,t/s}(t)$ for the tip/substrate can be expressed by the Langreth rules as
\begin{align}
    \hat J_\mathrm{emb,t/s}(t) =  \int \mathrm{d}s\left[\hat \Sigma_\mathrm{t/s}^R(t, s) \hat G^<(s,t) + \hat \Sigma_\mathrm{t/s}^<(t, s)\hat G^A(s,t)\right], 
\end{align}
with the advanced Green's function $\hat G^A_{\mathrm{t,s}}(t,t')$, and the retarded/lesser components of the self-energy $\hat \Sigma^{<,R}_\mathrm{t,s}(t,t')$ \cite{gianluca_stefanucci_robert_van_leeuwen_nonequilibrium_2013}.
The total charge in the sample is given by 
\begin{align}
    Q(t) = \mathrm{i}\mathrm{eTr}\left[\hat G^<(t,t)\right]
\end{align}
with $\mathrm{Tr[\cdot]}$ denoting the trace operation.

Hence, the total charge in the sample obeys
\begin{align}
    \frac{\mathrm{d}Q(t)}{\mathrm{d}t} + 2\mathrm{e}\sum_{\alpha = \mathrm{t,s}}\mathrm{ReTr}[\hat J_\mathrm{emb,\alpha}(t)] \label{eq:I_total} = 0.
\end{align}
Eq.~\ref{eq:I_total} constitutes a continuity equation for the charge transfer from the tip and the substrate to the sample. 
In particular, the tunneling current between tip and sample is given by
\begin{align}
    I(t) = -2\mathrm{eReTr}[\hat J_\mathrm{emb,t}(t)]
\end{align}
Note, that the unitary part $[\hat h(t),\hat G^<(t, t)]$ in Eq.~\ref{eq:dG<dt}  drops out upon the trace operation as the charge gains and losses in the sample cannot be captured in the unitary dynamics. 

The embedding self-energies are approximated in the wide band limit approximation (WBLA), modeling reservoirs that are in thermal equilibrium and characterized by featureless spectral functions with infinite bandwidths.
The retarded and lesser parts of the self-energies read
\begin{align}
    \hat \Sigma^R_\mathrm{s}(t-t') &= -\mathrm{i}\frac{\Gamma_\mathrm{s}}{2}\delta(t-t') \\
    \hat \Sigma^<_\mathrm{s}(t-t') &= \mathrm{i}\Gamma_\mathrm{s} \int \frac{\mathrm{d\omega}}{2\pi} f(\hbar\omega - \mu_\mathrm{s})e^{-\mathrm{i}\omega(t-t')} \label{eq:sigmasubstrate}
\end{align}
for the substrate, and
\begin{align}
    \hat \Sigma^R_\mathrm{t}(t-t') &= -\mathrm{i}\frac{\Gamma_\mathrm{t}}{2}\delta(t-t') \ket{\vec{r}}\bra{\vec{r}}  \\
    \hat \Sigma^<_\mathrm{t}(t,t') &= \mathrm{i}\Gamma_\mathrm{t} \int \frac{\mathrm{d\omega}}{2\pi} f(\hbar\omega - \mu_\mathrm{t} )e^{-\mathrm{i}\omega(t-t') - \mathrm{i}\Phi(t,t')}\ket{\vec{r}}\bra{\vec{r}} \label{eq:sigmatip}
\end{align}
for the tip.
Here, $f(\varepsilon) = 1/(1 + \exp(\beta \varepsilon))$ is the Fermi function with an inverse temperature $\beta$. 

Different chemical potentials $\mu_\mathrm{t/s}$ for tip and substrate are used to describe a static bias $\mu_\mathrm{s} - \mu_\mathrm{t} = \mathrm{e}V_\mathrm{DC}$. 
The bias modulation from the probe pulse is included in the phase of the tip self-energy, $\Phi(t,t') =  \frac{\mathrm{e}}{\hbar}\int_{t'}^t\mathrm{d}s V_\mathrm{THz}(s)$, and can be understood as a Peierls phase dressing the hopping integral between tip and sample.
The tunneling current thereby depends, via the phase $\Phi$, on the integral of the bias over earlier times. 
Hence, $I(t) \neq I(V_\mathrm{THz}(t))$ in our formalism, challenging the often employed description of ultrafast STM measurements where the current $I(t) = I(V_\mathrm{THz}(t))$ is simply a function of the instantaneous bias.

Note that $\hat{\Sigma}_\mathrm{s}(t,t')$ is proportional to the identity matrix and thereby couples equally to all sites of the sample. 
In contrast, the tip couples locally to a single lattice site at position $\vec{r}$, such that $\hat \Sigma_\mathrm{t}$ is proportional to the projector $\ket{\vec{r}}\bra{\vec{r}}$.

Due to the spatial locality of the tip self-energy, real-space simulations require separate computations for each tip position $\vec{r}$. 
Here, we assume that the coupling of the sample to the substrate is significantly stronger than to the tip.
Under this assumption we can evaluate $\hat J_\mathrm{emb,t}(t)$ for all tip positions simultaneously, but neglect it in the time evolution of $\hat G^<(t,t')$. 
This approximation implies that the sample will equilibrate to the substrate chemical potential $\mu_\mathrm{s}$ after an excitation. 
Consequently, the tunneling rate $\Gamma_\mathrm{t}$ only enters as a global prefactor in the calculation of the currents. 
Further details of the implementation can be found in Appendix~\ref{appx:A}.

The WBLA allows for a time-linear implementation that directly propagates the equal-time lesser component $\hat G^<(t,t)$ and the embedding terms $\hat{J}_\mathrm{emb,t/s}(t)$ \cite{tuovinen_time-linear_2023}.
\newline
\subsubsection{Floquet tunneling formalism}
As an instructive limiting case, we consider a continuous-wave (CW) drive modeled by the time-periodic Hamiltonian $\hat h(t) = \hat{h}(t + T)$ and ignore the time dependence of the bias by setting $V_\mathrm{THz}(t) = 0$. 
The reservoir coupling allows the heat from the drive to be dissipated such that a Floquet steady state is attained \cite{aoki_nonequilibrium_2014,dehghani_dissipative_2014, seetharam_controlled_2015,esin_quantized_2018, seetharam_steady_2019, park_steady_2022, mori_floquet_2023, liu_signatures_2025}.
This state can be modeled by employing the Floquet Green's function formalism.
Here, two-time functions $\hat{O}(t,t')$, such as Green's functions and self-energies, are represented by matrices in the enlarged Floquet replica space via a double Fourier transform with respect to the relative time $\tau = t- t'$ and the average time $\bar{t} = (t + t')/2 $ \cite{aoki_nonequilibrium_2014, liu_keldysh_2017, schuler_probing_2022}:
\begin{widetext}
\begin{align}
    \mathbf{O}_{mn}(\varepsilon) =\frac{1}{T}\int_0^T \mathrm{d}\bar{t}\int_{-\infty}^\infty \mathrm{d}\tau\  \hat O\left(\bar{t }+ \frac{\tau}{2},\bar{t }- \frac{\tau}{2}\right) e^{\mathrm{i}(m-n)\Omega \bar{t} + \mathrm{i}(\frac{\varepsilon}{\hbar} + \frac{m+ n}{2}\Omega )\tau  }, \label{eq:FloqG}
\end{align}    
\end{widetext}
with $\Omega = 2\pi/T$, and $m$ and $n$ enumerating the replica blocks.

In this Floquet representation, the different components $\mathbf{G}^{R/A/<}(\varepsilon)$ evaluate to 
\begin{align}
    \mathbf{G}^{R/A}(\varepsilon) = \left[\varepsilon - \mathbf{h} - \mathbf{\Sigma}^{R/A}(\varepsilon)\right]^{-1}\\
    \mathbf{G}^<(\varepsilon) =  \left[\mathbf{G}^{R}\cdot\mathbf{\Sigma}^<\cdot\mathbf{G}^A\right](\varepsilon).
\end{align}
Here $\mathbf{h}$ is the Hamiltonian in replica space defined by
\begin{align}
    \mathbf{h}_{mn} = \frac{1}{T}\int_0^T \mathrm{d}t e^{\mathrm{i}(m-n)\Omega t} \hat{h}(t) - m\hbar\Omega\delta_{mn},
\end{align} 
and $\mathbf{\Sigma}^{R/A/<}(\varepsilon)$ are the Floquet self-energies.
The convolutions along the Keldysh contour are replaced by matrix products to evaluate the cycle-averaged tunneling currents
\begin{align}
    \begin{split}
\bar{I} &= \int_0^T \frac{\mathrm{d}t}{T} I(t) \\
    &= -\frac{2\mathrm{e}}{\hbar}\int \frac{\mathrm{d}\varepsilon}{2\pi} \mathrm{ReTr}\left[ \mathbf{\Sigma}^R_\mathrm{t} \cdot\mathbf{ G}^< + \mathbf{\Sigma}^<_\mathrm{t} \cdot\mathbf{ G}^A \right]_{00}(\varepsilon)\\
    &= -\frac{\mathrm{e}\Gamma_\mathrm{t}}{\hbar} \int \mathrm{d}\varepsilon [ \bar {\mathcal{A}}_{\vec{r}}^<{(\varepsilon)} - f(\varepsilon - \mu_\mathrm{t})  \bar {\mathcal{A}}_{\vec{r}}{(\varepsilon)}   ] \label{eq:Ibar}.        
    \end{split}
\end{align}
In the second line, we inserted the Floquet representation of the tip embedding self-energy in the WBLA Eq.~\ref{eq:sigmatip}. 
The tunneling current can then be expressed in terms of the time-averaged LDOS $\bar{\mathcal{A}}_{\vec{r}}(\varepsilon)$ and the occupied LDOS $\bar{\mathcal{A}}_{\vec{r}}^<(\varepsilon)$, which are defined via the retarded and lesser components of the Floquet Green's function, respectively, as
\begin{align}
    \bar{\mathcal{A}}_{\vec{r}}(\varepsilon) = -\frac{1}{\pi}\mathrm{ImTr}\bra{\vec{r}}\mathbf{G}_{00}^R(\varepsilon)\ket{\vec{r}},\\
    \bar{\mathcal{A}}^<_{\vec{r}}(\varepsilon) = \frac{1}{2\pi}\mathrm{ImTr}\bra{\vec{r}}\mathbf{G}_{00}^<(\varepsilon)\ket{\vec{r}}.
\end{align}
We note the similarity of that result to the well-known relation of the differential conductance in Eq.~\ref{eq:dIdV}. 
It is evident that the time-averaged tunneling current can be used as a probe of the Floquet LDOS in close analogy to conventional STM. 

\subsection{Floquet engineering of band topology} \label{sec:Floq_form}
In this section we give a brief overview of topological Floquet engineering in graphene under a circularly polarized laser drive. 
The topological signature can be well understood based on the Floquet Bloch Hamiltonian whose quasienergy spectrum determines the Floquet Greens function and thus directly governs the time-averaged LDOS and the tunneling current as explained in Sec. \ref{sec:Floq_form}.

We start from the Bloch Hamiltonian for monolayer graphene
\begin{align}
    \hat h(\vec{k} ) = -\gamma_0 \begin{bmatrix}
        0 & f (\vec{k} )\\
        f^*(\vec{k}) & 0 
    \end{bmatrix} , \label{eq:TB_k_space}
\end{align}
where $f(\vec{k}) = e^{\mathrm{i} a\frac{k_y}{\sqrt{3}}} + 2 e^{-\mathrm{i}a \frac{k_y}{2\sqrt{3}}} \cos\left(\frac{ak_x}{2}\right)$, with the lattice constant of graphene of $a = 2.46~\mathrm{Å}$ and $\gamma_0 = 3.03~\mathrm{eV}$ \cite{mccann_electronic_2013}.
The Hamiltonian inherits its time dependence from the electromagnetic field by virtue of the Peierls substitution $\hat h(\vec k, t) \equiv \hat h(\vec k + \frac{\mathrm e}{\hbar} \vec A(t))$.
The vector potential $\vec{A}$ relates to the electric field of the laser via $\vec E = -\partial_t \vec{A}$ and is considered spatially homogeneous within the dipole approximation.
We further assume a circularly polarized drive $\vec{E}(t) = E_0 [\mathrm{cos}(\Omega t), \sin(\Omega t)]^\mathrm{T}$. 
In order to develop a basic understanding of how this setting gives rise to topological bands, we compute the Floquet bands close to the $K$ points.
Here, the bands are described by the Dirac Hamiltonian $\hat h = v_\mathrm{F}(p_x \sigma_{x} +\xi p_y \sigma_{y} )$ for $\xi = \pm $  at $K/K'$. 
$\sigma_{x/y}$ are Pauli matrices, $\vec p = (p_x, p_y)$ measures the crystal momentum relative to the $K$ points, and $v_\mathrm{F} = \sqrt{3}a\gamma_0/(2\hbar)$ is the Fermi velocity. 

In the weak-driving regime defined by
\begin{align}
    \eta = \frac{\mathrm{e}A_0 v_\mathrm{F}}{\hbar\Omega} \ll 1,
\end{align}
the Floquet Hamiltonian can be truncated to the replica sectors $m=0,\pm1$ \cite{usaj_irradiated_2014}. Here, $\vec p=\hbar(\vec k-\vec K_\xi)$ denotes the physical momentum measured relative to the valley $\vec K_\xi$, and $A_0=E_0/\Omega$ is the amplitude of the vector potential. The resulting truncated Floquet Hamiltonian is

\begin{widetext}
\begin{align}
\mathbf{h} =
\begin{bmatrix}
v_\mathrm{F}(p_x\sigma_x+\xi p_y\sigma_y)+\hbar\Omega
&
\displaystyle\frac{\mathrm{e}A_0v_\mathrm{F}}{2}
(\mathrm{i}\sigma_x+\xi\sigma_y)
&
0
\\[0.6em]
\displaystyle\frac{\mathrm{e}A_0v_\mathrm{F}}{2}
(-\mathrm{i}\sigma_x+\xi\sigma_y)
&
v_\mathrm{F}(p_x\sigma_x+\xi p_y\sigma_y)
&
\displaystyle\frac{\mathrm{e}A_0v_\mathrm{F}}{2}
(\mathrm{i}\sigma_x+\xi\sigma_y)
\\[0.6em]
0
&
\displaystyle\frac{\mathrm{e}A_0v_\mathrm{F}}{2}
(-\mathrm{i}\sigma_x+\xi\sigma_y)
&
v_\mathrm{F}(p_x\sigma_x+\xi p_y\sigma_y)-\hbar\Omega
\end{bmatrix}.
\end{align}

\end{widetext}

Downfolding onto the zeroth Floquet sector yields, to leading order in $\lvert\vec p\rvert$ and to second order in the drive amplitude,

\begin{align}
\hat h_\mathrm{eff}
&=
v_\mathrm{F}(p_x\sigma_x+\xi p_y\sigma_y)
-
\xi\frac{(\mathrm{e}A_0v_\mathrm{F})^2}{\hbar\Omega}\sigma_z.
\end{align}
The resulting mass term changes sign between the two valleys and opens a quasienergy gap
\begin{align}
\Delta_1
=
2\frac{(\mathrm{e}A_0v_\mathrm{F})^2}{\hbar\Omega}
=
2\eta^2\hbar\Omega
\end{align}
around $\varepsilon=0$ \cite{kitagawa_transport_2011}. This mass term is due to broken time-reversal symmetry as a consequence of the circularly polarized drive and realizes the low-energy structure of the Haldane model \cite{haldane_model_1988,hasan_colloquium_2010}, whose bands carry nonzero Chern numbers and support a quantum anomalous Hall response. Similarly, at the boundaries of the first Floquet Brillouin zone, $\varepsilon=\pm\hbar\Omega/2$, resonant hybridization between adjacent Floquet replicas opens a topological gap
\begin{align}
\Delta_2
=
\mathrm{e}A_0v_\mathrm{F}
=
\eta\hbar\Omega
\end{align}
in the weak-driving limit \cite{usaj_irradiated_2014}.

Away from the limiting case of weak driving, the perturbative analysis breaks down. 
Yet, the bands retain their topological nature, which only relies on the breaking of time-reversal symmetry by the circular drive. 
At increased field strengths, the hybridization with higher replica bands triggers further topological transitions that restructure the Floquet quasienergy spectrum including the topological gaps \cite{sentef_theory_2015}. 

In a finite geometry, the gaps are bridged by chiral edge states confined to the sample boundary, which disperse with opposite group velocities at opposite edges of the sample.
Thanks to their topological protection, the edge states are robust to backscattering at impurities \cite{perez-piskunow_floquet_2014}. 
The chirality of the edge states is inverted when switching between right-handed and left-handed circularly polarized driving \cite{oka_photovoltaic_2009}. 

\section{Results}
\label{sec:results}
Having established the conceptual framework, we now turn to the question of how Floquet signatures can be accessed under ultrafast driving using THz-STM. 
Specifically, we will show that THz-STM measurements are able to capture all of the relevant signatures of the Floquet topological state, namely the opening of dynamical gaps with their characteristic dependence on drive parameters, the formation of chiral edge states, as well as their impurity scattering behavior. 

\subsection{Bulk gap openings} \label{sec:bulkgaps}
In Fig.~\ref{fig:1-Floquet_bands+DoS} b) we show the Floquet band structure of graphene around the $K$ point for a circularly polarized electric field.
Here, the color code indicates the spectral weight $\bar{\mathcal{A}}_{\vec{k}}(\varepsilon) = -\frac{1}{\pi}\mathrm{ImTr}\bra{\vec{k}}\mathbf{G}_{00}^R(\varepsilon)\ket{\vec{k}}$, which highlights the time-averaged spectral weight of the Floquet sidebands. 
Furthermore, we show the Floquet LDOS in bulk in Fig.~\ref{fig:1-Floquet_bands+DoS} c). 
Both the Haldane gap $\Delta_1$ and the Floquet hybridization gap $\Delta_2$ are clearly visible in the band structure. The former is smeared in the LDOS plot due to a finite broadening $\Gamma = 10~\mathrm{meV}$ representing the finite coupling to the environment. 

\begin{figure*}
    \centering
\includegraphics[width=\linewidth]{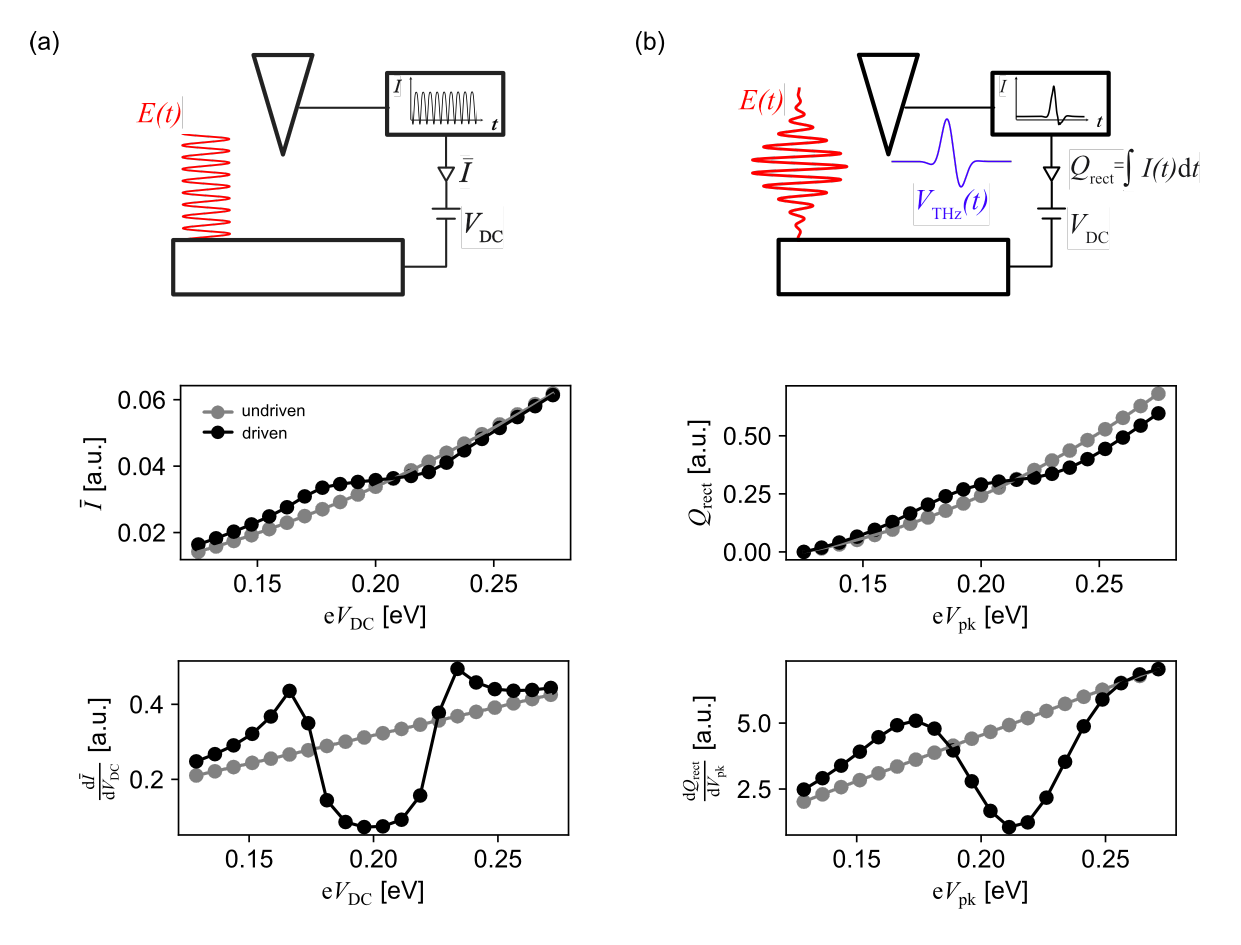}
    \caption{Measurement protocols for lightwave-driven tunneling spectroscopy of Floquet states. 
  (a) Idealized configuration with a continuous-wave (CW) drive of amplitude $E_0$ inducing a Floquet steady state. The time-averaged current $\bar{I}$ is measured as a function of the static bias $V_\mathrm{DC}$. 
  (b) Experimental configuration with a multi-cycle driving pulse with peak field strength $E_0$ inducing the transient Floquet state. The DC bias is modulated by a single-cycle THz probe pulse $V_\mathrm{THz}$.
  The rectified current $Q_\mathrm{rect}$ is recorded as a function of the peak THz bias $V_{\mathrm{pk}}$.
  The upper panels of a) and b) illustrate the measurement protocols; the middle and lower panels show $\bar{I}$ [$Q_\mathrm{rect}]$ and tunneling conductances $\mathrm{d} \bar{I}/\mathrm{d}V_{\mathrm{DC}} $ [$\mathrm{d}Q_\mathrm{rect}/\mathrm{d}V_{\mathrm{pk}}$] as a function of $V_\mathrm{DC}$ [$V_\mathrm{pk}$] for driven and undriven cases, respectively.
  Dynamical band gaps appear as minima in the tunneling conductances.  
  Simulations were performed for $E = 350~\mathrm{kV\,cm^{-1}}$ and $\hbar\Omega = 0.4~\mathrm{eV}$. }
    \label{fig:3-bulkgap}
\end{figure*}

To demonstrate the applicability of our formalism for studying Floquet states by STM, we first model STS-type differential conductance spectra of the light-induced bulk gaps. 
In Fig.~\ref{fig:3-bulkgap} we compare Floquet steady-state calculations of the tunneling current with the real-time tunneling results using ultrafast driving and probing. 
These two scenarios can be associated with two distinct experimental protocols as illustrated in the respective upper panels.
\revise{A key point of the following discussion is that the two protocols provide closely related spectroscopic information. In the limit of sufficiently long pump and probe pulses, where the pulse envelopes vary slowly compared to the intrinsic relaxation times of the material, the full real-time description approaches the Floquet steady-state picture. This correspondence is well established for complementary probes of Floquet states, such as time-resolved photoemission spectroscopy \cite{schuler_how_2020,schuler_probing_2022}, and provides a useful benchmark for theoretical simulations as well as a guide for interpreting experimental data. At the same time, the real-time formulation remains essential whenever transient carrier populations, interband transitions, or finite-pulse effects play an important role.}

In the steady-state Floquet formalism in Fig.~\ref{fig:3-bulkgap} a), we assume a CW drive with time-periodic electric field $E(t)$, and only a static DC bias $V_\mathrm{DC}$ is applied to control the steady-state occupations of the sample.
This configuration primarily serves as an instructive limiting case for the more realistic pump-probe protocol. 
\revise{In addition it also defines an optical STM set-up, as recently realized in Ref.~\onlinecite{wu_optical_2024}, which is distinct from THz-STM. }

The oscillations of the tunneling current $I(t)$ cannot be resolved in the experiment, and the output measured at the STM preamplifier corresponds to the time-averaged current $\bar{I}$ given by Eq.~\ref{eq:Ibar}. 

The configuration shown in Fig.~\ref{fig:3-bulkgap} b) represents an idealized version of an actual pump-probe THz-STM measurement. 
Both the excitation pump pulse and the THz probe pulse now have a finite temporal width. 
The time-dependent bias $V_\mathrm{THz}(t)$ is added to the static DC bias $V_\mathrm{DC}$ and the experimental observable is the rectified charge $Q_\mathrm{rect}$ defined in Eq. \ref{eq:q_rect}.
If the width of the THz probe pulse exceeds several periods, $Q_\mathrm{rect}$ is dominated by the cycle-averaged current enabling a closer comparison to the steady-state $\bar{I}$.

As an example, we simulate  STS-type differential conductance measurements of the Floquet hybridization gap at $\hbar\Omega/2$. \revise{We focus on the hybridization gap because it is parametrically larger than the Dirac-point Haldane gap in the weak-driving regime and therefore provides a particularly clear benchmark for the finite-pulse tunneling protocol. The latter gap remains an important target for local spectroscopy in situations where it competes with static sublattice-symmetry-breaking masses.}

We use a pump field
\begin{align}
    \vec {E}(t) =E_0 \begin{pmatrix}
        \cos(\Omega t) \\
        \sin(\Omega t)
    \end{pmatrix} \times \left\{\begin{array}{ll} 1 & \mathrm{Floquet}, \\
         e^{-\frac{t^2}{2\sigma_\mathrm{pump}^2} } &  \mathrm{real-time},\end{array}\right. \label{eq:pump}
\end{align}
with $E_0 = 350~\mathrm{kVcm}^{-1}$, $\hbar \Omega = 0.4~\mathrm{eV}$, and $\sigma_\mathrm{pump} = 240~\mathrm{fs}$ for the pulsed excitation.
In the steady-state Floquet simulation, the chemical potential of the tip is kept at $\mu_\mathrm{t} = 0$, while the substrate chemical potential at $\mu_\mathrm{s}$ is modified to simulate a DC bias sweep with $\mathrm{e}V_\mathrm{DC} = \mu_\mathrm{s} - \mu_\mathrm{t}$.

For the real-time simulation, i.e. THz-STS of transient Floquet states, we include a THz bias pulse as probe of the form 
\begin{align}
    V_\mathrm{THz}(t) = V_0\cdot \cos(2\pi \nu (t- t_\mathrm{probe}) + \varphi_\mathrm{CEP})e^{-\frac{(t- t_\mathrm{probe})^2}{2\sigma_\mathrm{probe}^2} }, \label{eq:probe}
\end{align}
using $\sigma_\mathrm{probe} = 80 ~\mathrm{fs}$, $ \nu = 3 ~\mathrm{THz}$, and a probe time delay of $t_\mathrm{probe} = 0$.  The carrier envelope phase is $\varphi_\mathrm{CEP} = 2\pi/3 $.
The amplitude of the probe pulse is scaled by $V_0$ and coincides with $V_0$ for $\varphi_\mathrm{CEP} = 0$.
A static bias is further included by setting $\mu_\mathrm{t} = 0.125~ \mathrm{eV}$ and $\mu_\mathrm{s} = 0$. 

This combination of DC bias and THz bias is chosen here to set the energy window sampled by the THz probe pulse close to the Floquet band gap at $\varepsilon = 0.2~ \mathrm{eV}$ in order to highlight the Floquet-induced gap in the simulated spectra.
In an actual experiment, one needs to consider that a large $V_\mathrm{DC}$ will lead to large undesired static tunneling currents, which increase noise and limit the range over which $V_\mathrm{DC}$ can realistically be swept. 
The implementation of advanced probing schemes, such as the combination of two probe pulses from which one replaces $V_\mathrm{DC}$, may provide a solution. 

For the simulations in Fig.~\ref{fig:3-bulkgap}, we set $\Gamma_\mathrm{s} = 10~ \mathrm{meV}$, $\Gamma_\mathrm{t} = 0.2~ \mathrm{meV}$ and choose the temperature of tip and substrate to be $T = 8$ K. 
 
The middle and bottom panels in Figures \ref{fig:3-bulkgap} a) and \ref{fig:3-bulkgap} b) show the results for the hybridization bulk gap in graphene. 
The grey and black lines show the data with and without the Floquet drive, respectively. 
In the steady-state Floquet scenario (Fig.~\ref{fig:3-bulkgap} a)), we plot $\bar{I}$ as a function of $V_\mathrm{DC}$, while in the ultrafast driving case, $Q_\mathrm{rect}$ is plotted as a function of $V_\mathrm{pk} = \max[V_\mathrm{DC} + V_\mathrm{THz}(t)]$ which is controlled by sweeping $V_0$.

The band gap can be estimated from the perturbative analysis to be $\Delta_2 = 0.056~\mathrm{eV}$. Its emergence gives rise to a plateau in the generalized current-voltage characteristics centered near the bias $\mathrm{e}V_\mathrm{DC/pk}=\hbar\Omega/2=0.2~ \mathrm{eV}$ (middle panels). 
To examine the spectroscopic capabilities of STM to characterize such light-induced Floquet gaps, we evaluate the numerical derivatives $\mathrm{d} \bar{I}/\mathrm{d}V_{\mathrm{DC}} $ and $\mathrm{d}Q_\mathrm{rect}/\mathrm{d}V_{\mathrm{pk}}$ (bottom panels). In both cases, the band gap is visible as a pronounced dip in the generalized differential conductance curves. For the steady-state case of CW driving, the Floquet tunneling conductance follows the LDOS shown in Fig.~\ref{fig:1-Floquet_bands+DoS} c) around the band gap.
Experimentally, the achievable energy resolution in this case would be set by the bias modulation amplitude used for recording STS at sufficiently low temperature. 

In the real-time case of THz-STS, plotted in Fig.\ref{fig:3-bulkgap}(b), the spectral profile around the band gap is slightly washed out compared to the steady-state case. 
Yet, the $\mathrm{d}Q_\mathrm{rect}/\mathrm{d}V_{\mathrm{pk}}$ spectra still resemble the Floquet simulation results, and the hybridization gap can be well resolved.
In THz-STS, the energy resolution is set by the peak amplitude of the applied THz bias, which defines the energy window over which $Q_\mathrm{rect}$ will be integrated. For the ultrafast real-time calculations, our simulations directly account for the finite energy resolution of the probe.
In addition, the different shapes of the gaps observed in the tunneling spectra are explained by the integration over the oscillatory single-cycle THz bias waveform. Specifically, sweeping $V_\mathrm{pk}$, either via $V_\mathrm{0}$  or $V_\mathrm{DC}$, will not only change the contribution of states around the maximum of the bias pulse, but over the whole bias range covered by [$   V_\mathrm{DC}+   V_\mathrm{THz}(t)]$.
In particular, the negative half-cycle of the THz bias also scales with $V_\mathrm{0}$.
In the example in Fig. \ref{fig:3-bulkgap} b),  the instantaneous tunneling current originating from the positive half-cycle becomes suppressed when it enters the hybridization gap, while the negative half-cycle still sweeps across below-gap states at lower energies. This explains the shift of the minimum of the $\mathrm{d}Q_\mathrm{rect}/\mathrm{d}V_{\mathrm{pk}}$ spectra to higher values along $V_\mathrm{pk}$. \revise{The tunneling conductance is further distorted by the memory effect of the tunneling junction. Consequently, the real-time THz-STS signal provides clear evidence for the existence of the Floquet gap, but the apparent minimum in $\mathrm{d}Q_\mathrm{rect}/\mathrm{d}V_{\mathrm{pk}}$ should not be identified directly with the exact gap position. A quantitative extraction of gap energies therefore requires comparison with realistic simulations and careful calibration of the THz bias waveform. This limitation does not undermine the applicability of the approach, since the absolute field strength inside the junction is generally not known independently and must in any case be inferred from calibration.}

The results in Fig.\ref{fig:3-bulkgap} are a specific example focusing on the opening of the dynamic gap at $\hbar\Omega/2$. It is straightforward to extend the presented scheme to arbitrary DC biases and THz bias waveforms. More broadly, the formalism is not restricted to study the Floquet LDOS in graphene, but provides a versatile framework for probing the nonequilibrium spectral features that emerge in the LDOS of driven quantum materials. 

\subsection{Edge state spectroscopy in graphene nanoribbons}
\label{sec:ribbons}

\begin{figure*}
    \centering
\includegraphics[width=\linewidth]{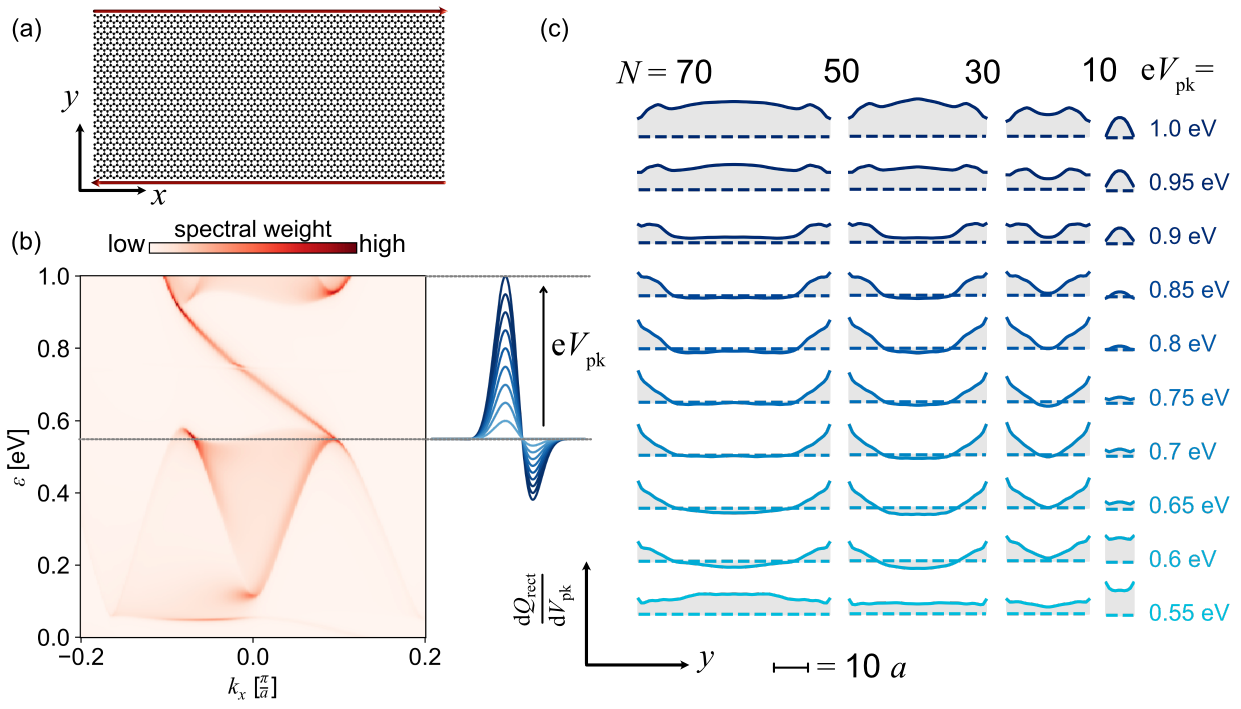}
    \caption{\revise{ Floquet edge states in a graphene nanoribbon.
        (a) Real-space schematic: Zigzag ribbon with counter-propagating edge modes as indicated by red arrows.
        (b) Quasienergy spectrum of a semi-infinite graphene ribbon with zigzag edges along the $x$ direction, projected onto the first 30 unit cells measured from the edge.
        The driving parameters are $E_0 = 10~\mathrm{MV\,cm^{-1}}$ and $\hbar\Omega = 1.5~\mathrm{eV}$, which are chosen to create large enough gap to ensure sufficient real-space localization within our computational limits.
        The edge state bridges the dynamical gap at $\varepsilon = 0.75~\mathrm{eV}$.
        The blue curves shows the THz probe pulse with variable amplitude as scanned for the simulations in (c).
        (c) Simulated generalized differential conductance $\mathrm{d}Q_\mathrm{rect}/\mathrm{d}V_\mathrm{pk}$ along the $y$ direction for the same driving parameters as in (b) and varying ribbon width of $N = 70, 50, 30, 10$ unit cells. The $\mathrm{d}Q_\mathrm{rect}/\mathrm{d}V_\mathrm{pk}$ values are obtained by sweeping $V_\mathrm{pk}$ and numerical differentiation of the $Q_\mathrm{rect}-V_\mathrm{pk} $  curve.
        The profiles are vertically offset for clarity and referenced to the zero level defined by $\mathrm{d}Q_\mathrm{rect}/\mathrm{d}V_\mathrm{pk}=0$ (dashed line).
        Upon sweeping $V_\mathrm{pk}$ through the bulk gap, the edge states manifest as peaks at the ribbon edges, which decays towards the bulk for ribbon sizes down to $N = 30$ unit cells.}}
    \label{fig:4-edgestates}
\end{figure*}

Having established pump-probe tunneling spectroscopy for bulk states, we now turn to study the topological edge states in graphene. 
To this end, we apply our formalism to a tight-binding model of graphene nanoribbons with zigzag edges.
Via a Fourier transform of Eq.~\ref{eq:TB_k_space} with respect to $k_y$, the hopping matrix elements between the one-dimensional carbon zigzag labeled by chains with indices $i$ and $j$ are obtained as
\begin{align}
\begin{split}
v_{ij} &=  \cos\left(\frac{a}{2}\left(k_x + \frac{\mathrm{e}}{\hbar}A_x\right)  \right)\begin{bmatrix}
        0 &e^{-\mathrm{i}\frac{\mathrm{e} aA_y}{2\sqrt{3}\hbar}}\\
        e^{\mathrm{i}\frac{ \mathrm{e}a A_y}{2\sqrt{3}\hbar}} & 0
    \end{bmatrix} \delta_{ij} \\
    &+\begin{bmatrix}
        0 & 1\\
        0 & 0
    \end{bmatrix}e^{\mathrm{i}\frac{\mathrm{e} aA_y}{\sqrt{3}\hbar} }  \delta_{i, j-1 }  +  \begin{bmatrix}
        0 & 0\\
        1 & 0
    \end{bmatrix} e^{-\mathrm{i}\frac{\mathrm{e} aA_y}{\sqrt{3}\hbar} } \delta_{i, j+1 } .
\end{split}
\end{align}
The zigzag Hamiltonian is then given by
\begin{align}
    \hat{h}_\mathrm{zz} = -\gamma_0\sum_{i,j = 1}^{N} v_{ij} \ket{i}\bra{j} 
\end{align}
with a finite number $N$ of unit cells in the $y$ direction, while translational invariance is maintained along $x$. The ribbon width is given by $W = \frac{\sqrt3Na}{2} $.

The ribbon geometry is sketched in Fig.~\ref{fig:4-edgestates} a) with the counter-propagating edge states being indicated by red arrows. 
We note that, due to the topological nature of the edge states, the results will not depend on the specific shape of the edge as demonstrated in Ref.~\onlinecite{usaj_irradiated_2014}. 

Before discussing the results, we briefly comment on a practical difficulty encountered in real-space simulations of edge states.
The edge state wave functions decay over a length scale that increases with decreasing band gap. 
A perturbative estimate yields a decay length $\xi \sim 2\hbar v_\mathrm{F}/\Delta$ for the edge states lying in the hybridization gap in the perturbative weak-drive regime \cite{usaj_irradiated_2014}.
Typical driving parameters that can be used in experiments give rise to band gaps on the order of 10 meV, such that ribbons of a width of several 100 lattice constants are required in order to obtain properly localized and spatially separated edge states. 
As the real-time simulations become increasingly computationally demanding for increasing Hilbert space dimensions, we limit our simulations to a ribbon of up to $N = 70$. Therefore, to ensure sufficient spatial separation and to avoid hybridization of opposite edge states, we need to adapt the driving parameters to values that cannot be achieved in realistic experimental settings. In particular, we increase the photon energy to $\hbar\Omega = 1.5~\mathrm{eV}$ and the field strengths of $E_0 = 10~\mathrm{MVcm}^{-1} $. However, we note that this choice is only for practical reasons to keep the microscopic simulations feasible. 
\revise{While this choice changes the localization length and energy scales, it does not affect the basic mechanism: circular driving creates topological Floquet gaps that are bridged by chiral edge states, and these states can be detected through their spatially resolved tunneling signal.} 
Furthermore, in reality, exfoliated or substrate-grown graphene samples extend over micrometers, a scale on which edge states for realistic drive parameters will be well localized. Thus, the overall conclusions do not depend on the specific choice of the drive parameters.

The Floquet band structure for our driving parameters and for a semi-infinite ribbon projected to the first 30 unit cells is shown in Fig.~\ref{fig:4-edgestates} b).
Here, the bands for the semi-infinite geometry are numerically obtained via the recursive Green's function method. 
The Floquet hybridization gap centered at $\varepsilon = 0.75$ eV exceeds 400 meV and is bridged by a clearly discernible edge state.

We then repeat the pump-probe protocol discussed for the bulk case in the previous section, with the definition for the pump and probe pulses as in Eq.~\ref{eq:pump} and \ref{eq:probe} and the pulse parameters $\sigma_\mathrm{pump} = 150$ fs, $\sigma_\mathrm{probe} = 50$ fs, $\nu = 3$ THz, and $\varphi_\mathrm{CEP} = 2\pi/3$.
Tip and substrate enter at a temperature of $T = 16 ~\mathrm{K}$ and with tunneling rates $\Gamma_\mathrm{s} = 20 ~\mathrm{meV}$ and $\Gamma_\mathrm{t} = 0.4 ~\mathrm{meV}$, respectively.
A static bias $\mu_\mathrm{s} = 0$ eV and $\mu_\mathrm{t} = 0.55 $ eV is further applied, and the THz probe pulse amplitude is swept such that $V_\mathrm{pk}$ is scanned through the bulk gap, as indicated by the blue bias waveforms sketched in Fig.~\ref{fig:4-edgestates} b).
In  Fig.~\ref{fig:4-edgestates} c) we plot the spatial profile of the differential tunneling conductance at different energies $eV_\mathrm{pk}$ for different ribbon widths of $N= 70, 50, 30, 10$ unit cells. The values $\mathrm{d}Q_\mathrm{rect}/\mathrm{d}V_{\mathrm{pk}}$ are obtained by numerical differentiation of the spatially dependent $Q_\mathrm{rect}$-$V_\mathrm{pk}$-curves and are averaged over the two sublattices.

For the large ribbon ($N = 70$) and energies in the range $0.6~ \mathrm{eV} \lesssim V_\mathrm{pk} \lesssim 0.9~\mathrm{eV}$, the generalized differential conductance $\mathrm{d}Q_\mathrm{rect}/\mathrm{d}V_{\mathrm{pk}}$ is significantly enhanced at the edges and decays when scanning into the bulk. This provides a direct real-space tunneling signature of light-induced edge states localized at the ribbon boundaries. A localized edge state signal can be measured down to a ribbon widths of $N = 30$ unit cells. A systematic progression is observed as the width of the ribbon is reduced, and at $N = 10$ the edge states from the opposite edges hybridize and annihilate one another.

\subsection{Floquet quasiparticle interference}
\label{sec:qpi}

The spatially resolved results in Fig.~\ref{fig:4-edgestates} c) demonstrate the emergence of light-induced edge states under periodic driving. However, the temporally and spatially modulated conductance alone does not directly image the chiral edge mode dispersion. 

Quasiparticle interference (QPI) infers momentum-space information from real-space modulations of the LDOS generated by defects or boundaries and measured in STM. By Fourier transforming these standing-wave patterns, one identifies the dominant scattering wave vectors at a given energy and can reconstruct the underlying dispersion and scattering selection rules \cite{mallet_role_2012, avraham_quasiparticle_2018, yin_probing_2021}. In the following, we take this well-established static QPI concept as a starting point and extend it to periodically driven graphene in order to access the driven edge-mode dispersion from real-space STM observables. Since full real-space, real-time simulations are computationally demanding, we perform the Floquet QPI analysis in the steady-state Floquet framework which we have shown to capture the most relevant spectroscopic signatures in previous sections. Extending the construction of Floquet QPI patterns to the real-time setting is conceptually straightforward.

\begin{figure*}
    \centering
\includegraphics[width=\linewidth]{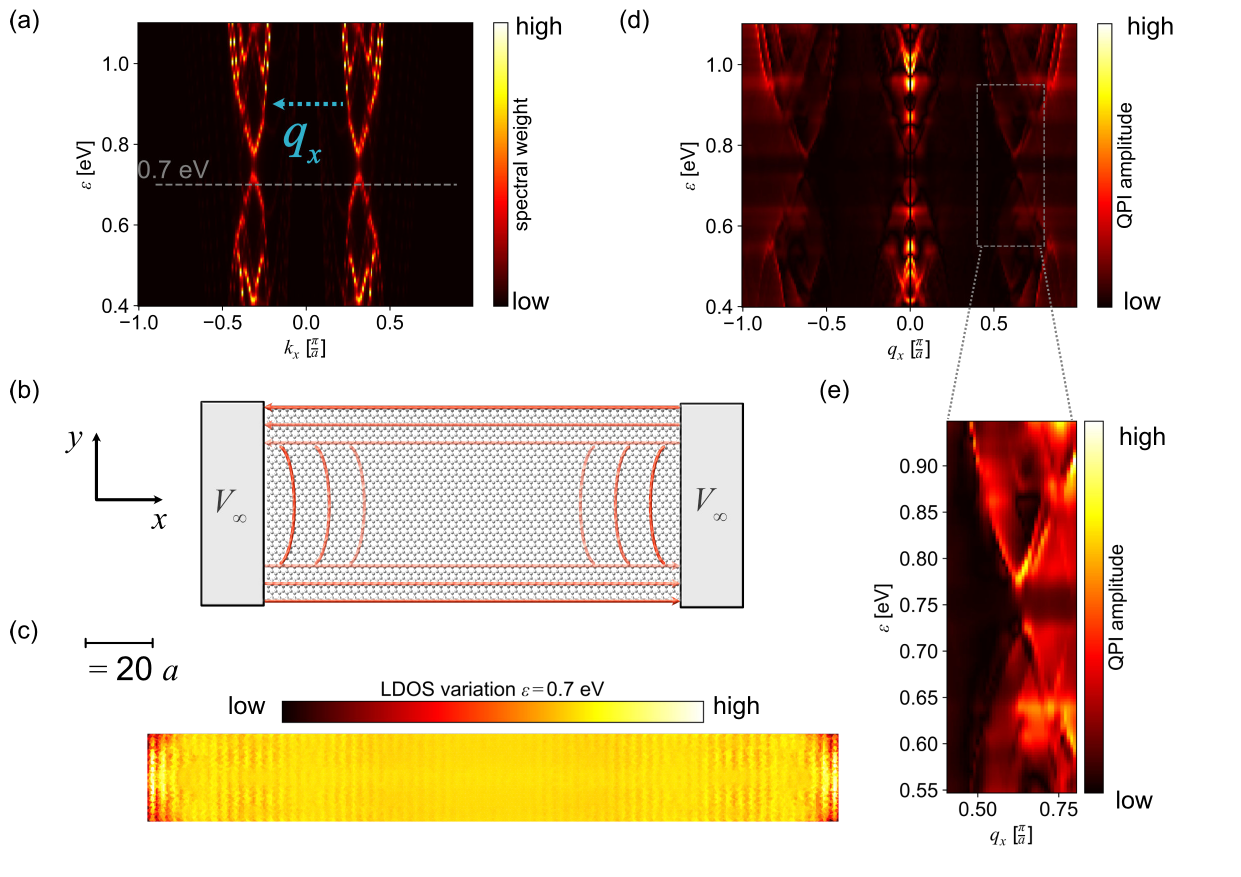}
    \caption{Edge state interferometry in a narrow nanoribbon. (a) Spectral function for $N = 30$ and $E_0 = 10~\mathrm{MVcm}^{-1}$, $\hbar \Omega = 1.5~\mathrm{eV}$.  Due to inter-edge coupling, two edge states are visible. The dashed line marks the energy used for the LDOS calculation in (c). (b) Sketch of the device architecture for edge state interferometry: A narrow ribbon allowing a finite hybridization of edge states in the bulk is restricted from both sides leading to Floquet QPI patterns. (c) LDOS variation due to hard-wall potential at $\varepsilon = 0.7 ~ \mathrm {eV}$. (d) QPI map as defined in the main text, showing the scattering channels that are allowed between the states plotted in (a). (e) Zoom into the QPI map region marked by the rectangle in (d), showing an image of the unperturbed ribbon band structure. The mean QPI intensity, corresponding to a peak at $q_x = 0$, is subtracted to improve clarity.}
    \label{fig:4-QPI}
\end{figure*}
Inspired by Ref.~\onlinecite{stuhler_effective_2022}, we propose to probe the edge state momentum-energy dispersion by inter-edge coupling in narrow nanoribbons or similar confined geometries. 
To this end, we again consider $E_0 = 10~\mathrm{MVcm}^{-1}$ and $\hbar\Omega = 1.5~\mathrm{eV}$ and choose the ribbon width along $y$ as $N =30$ being narrow enough to allow a finite hybridization of the edge states in the center region of the ribbon but still wide enough to prevent their annihilation.
The edge state stability is confirmed by the conductance profiles in Fig.~\ref{fig:4-edgestates}(c) and the momentum-resolved spectral function in Fig.~\ref{fig:4-QPI}(a), which show that the edge states remain intact over a broad spectral range.

Let us now discuss how to choose the ribbon geometry to enable Floquet QPI. We make use of the finite inter-edge coupling by simulating an interferometric device where the ribbon is restricted by a hard-wall potential of energetic height $V_\mathrm{\infty} = 10^7~\mathrm{eV}$ at the ends along the $x$ direction, see Fig.~\ref{fig:4-QPI} b).
Unlike well separated and thereby topologically protected edge states in wide ribbons, the coupled edge states can backscatter at the hard walls and thereby form standing wave interference patterns. 
This setup serves as a simplified model for more realistic experimental configurations. \revise{The proposed principle is not tied to the specific, computationally motivated drive parameters or ribbon dimensions used here. More realistic implementations could use larger graphene samples in which edge states remain well localized but acquire a controlled weak coupling in engineered slit, constriction, or interferometer geometries, as discussed in Ref.~\cite{stuhler_effective_2022}. In such devices, backscattering can be introduced, for example, by geometric kinks or terminations along the edge \cite{stuhler_effective_2022}.} Here the backscattering is for example induced by geometric kinks along the edge \cite{stuhler_effective_2022}.
We use a Floquet T-matrix formalism (see. Appendix \ref{appx:B}) to calculate the Floquet LDOS correction $\delta \bar{\mathcal{A}}_{\vec r}(\varepsilon)$ due to the hard walls.
The resulting spatial distribution is plotted in Fig.~\ref{fig:4-QPI} c) for an exemplary energy $\varepsilon=0.7$ eV, featuring a distinct quasiparticle interference pattern.

We then vary $\varepsilon$ and compute the energy-resolved QPI amplitude $\Pi_{ q_x}(\varepsilon) = \vert \langle \mathcal F_x [\delta \bar {\mathcal A}_{\vec r}(\varepsilon)]\rangle_y \vert$ as the Fourier transform of the standing-wave Floquet LDOS modulation with respect to $x$, averaged along $y$. 
The QPI maps in Fig.~\ref{fig:4-QPI} d) and the zoom-in Fig.~\ref{fig:4-QPI} e) show the scattering intensity  $\Pi_{ q_x}(\varepsilon) $ as a function of energy and momentum transfer $q_{ x}$. Clearly, for a given energy, $\Pi_{q_x}(\varepsilon)$ shows peaks at dominant scattering wave vectors (as illustrated by the example scattering vector $q_x$ in Fig.~\ref{fig:4-QPI} a)). 
In conventional QPI, these are typically wave vectors connecting states on equi-energy contours at energy $\varepsilon$ (or Fermi-surface nesting vectors for  $\varepsilon=E_\mathrm{F}$). For the one-dimensional ribbon band structure, the contours reduce to discrete momentum points, so that scattering occurs between momentum points rather than extended surfaces.

%The QPI map (Fig.~\ref{fig:4-QPI} d)) is dominated by intensity near , which do allow a straightforward interpretation at the given momentum resolution.
Pronounced peaks around $q_x = \pm 0.6~ \frac{\pi}{a}$ (Fig.~\ref{fig:4-QPI} e)) indicate strong inter-valley scattering. The Dirac-cone dispersion of the edge states, hybridized due to inter-edge coupling, is clearly imprinted on the Floquet QPI pattern. Floquet QPI thus provides a tool to image the nanoribbon's energy-momentum dispersion including chiral Floquet edge modes.
In addition, by tuning the inter-edge coupling via the ribbon width or the driving parameters and monitoring the gradual suppression of the QPI intensity, one could in principle obtain evidence of the absence of backscattering and the unidirectional nature of the isolated edge states.

\subsection{Circular pump dichroism in the presence of chiral impurities}
\label{sec:dichroism}
\begin{figure*}
    \centering
\includegraphics[width=\linewidth]{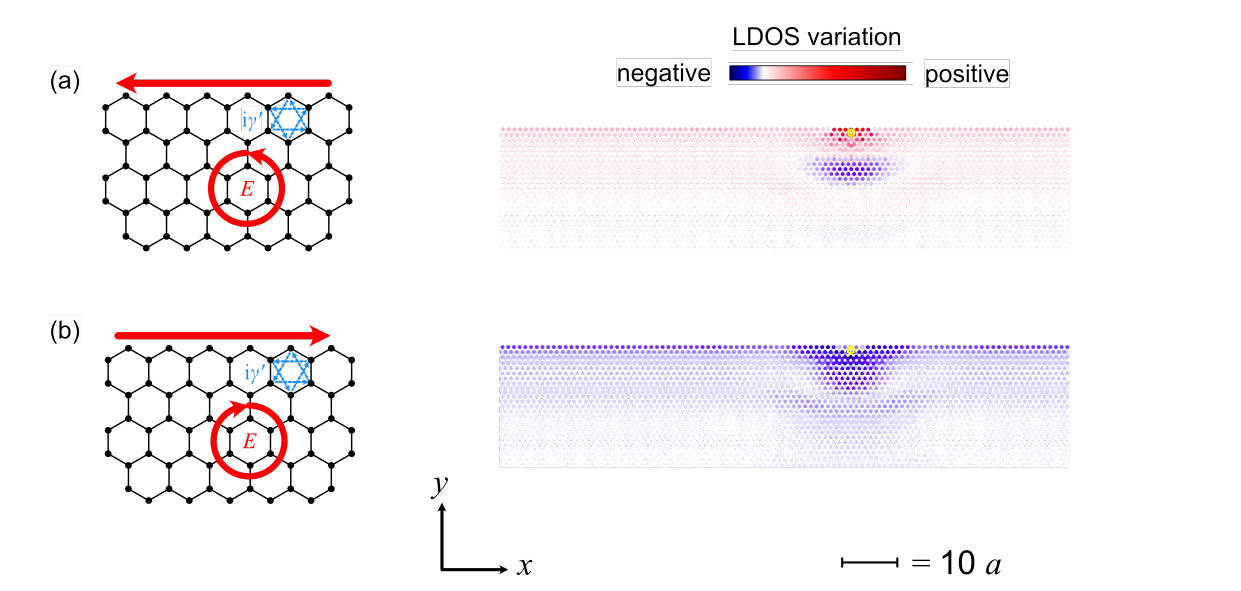}
    \caption{Polarization-dependent Floquet LDOS near a chiral defect. The chiral defect is implemented by Haldane-like complex next-nearest neighbor hopping amplitudes $\mathrm i \gamma'$ with $\gamma' = 0.7~\mathrm{eV}$ as illustrated in the real-space sketch of the graphene lattice (left panel). The right plots show the LDOS variation $\delta \bar{\mathcal{A}}_{\vec r}(\varepsilon)$ due to the presence of the defect (marked in yellow).  
     (a) and (b) highlight the difference upon switching from right- to left-handed circular polarization, leading to an enhancement or a suppression of the LDOS. The LDOS is computed for a semi-infinite ribbon with zigzag edges and driving parameters $E_0 = 10~\mathrm{MVcm^{-1}}$ and $\hbar \Omega = 1.5~\mathrm{eV}$ in the middle of the gap at $\varepsilon = 0.75~\mathrm{eV}$.}
    \label{fig:5-QPI}
\end{figure*}

We finally propose a measurement to directly probe the chiral nature of Floquet edge modes under circularly polarized irradiation. The core idea builds upon the fact that the edge state chirality switches sign when the helicity of the circular drive is reversed. 
Specifically, we examine the polarization dependence of the edge state LDOS in the presence of a chiral impurity that locally breaks time-reversal symmetry.

We model the chiral impurity by introducing Haldane-like complex next-nearest-neighbor hopping amplitudes $\mathrm{i}\gamma'$ \cite{haldane_model_1988} within a single edge-adjacent hexagon of the ribbon, choosing $\gamma'=0.7~\mathrm{eV}$. \revise{Here we do not aim at a microscopic description of a specific defect, but rather introduce a proof-of-principle perturbation that locally breaks time-reversal symmetry. Accordingly, $\gamma'$ is chosen comparable to the light-induced topological gap in order to produce a clearly resolvable local response. Physically, this local ``Berry-flux'' defect may be viewed as a minimal model for a nanoscale time-reversal-symmetry-breaking region that couples to the orbital motion of the charge carriers. Possible realizations include local magnetic textures generated by magnetic adatoms or intercalants, such as Co-intercalated graphene \cite{uchoa_localized_2008,decker_atomic-scale_2013}, or defect- and adatom-induced magnetic moments that can be tuned electrostatically \cite{nair_dual_2013}. More broadly, analogous physics could arise from the interplay between Floquet driving and intrinsic topological or valley-polarized electronic order, for example in regions carrying an anomalous-Hall, or Haldane, mass \cite{nandkishore_quantum_2010}.} 
In contrast to an extended region that would realize a full Haldane phase, we keep the perturbation strictly local so that it acts as a chiral impurity that locally breaks time reversal symmetry and competes with or reinforces the global Floquet mass depending on the drive helicity.
Because this effect is sensitive to the relative sign of the drive's helicity and the impurity's chirality, the LDOS around the impurity becomes circularly dichroic with respect to the polarization of the Floquet drive.

We compute the Floquet LDOS variation $\delta \bar{\mathcal{A}}_{\vec r}(\varepsilon)$ due to the impurity for a semi-infinite ribbon with $E_0 = 10~\mathrm{MVcm}^{-1}$ and a frequency $\hbar\Omega =1.5~\mathrm{eV} $ for an energy $\varepsilon = 0.75~\mathrm{eV}$ in the middle of the gap. For details see Appendix \ref{appx:B}. 
As anticipated, we find a difference between left- and right-handed circular polarization, which we compare in Fig.~\ref{fig:5-QPI}.
Depending on the relative sign of edge state and impurity chiralities, we observe either an enhancement (Fig.~\ref{fig:5-QPI} a)) or a suppression (Fig.~\ref{fig:5-QPI} b)) of the edge state LDOS.
We interpret this as a real-space signature of the enhancement or suppression (or even inversion) of the Floquet band gap, respectively. 
In real-space language, the effect of the chiral impurity is either to further confine the edge state or to delocalize and ultimately erase it.
From a semiclassical perspective, this renormalization can be viewed as a microscopic manifestation of the Aharonov-Bohm effect \cite{aharonov_significance_1959}.
The chiral edge state, while traveling unidirectionally, effectively splits around the impurity hexagon. 
The relative phase acquired from the local Berry flux leads to constructive or destructive interference of neighboring paths and thereby a spatially extended LDOS variation even away from the immediate impurity site.
\revise{Whereas the interpretation in terms of local variations of the Floquet gap and the semiclassical Aharonov-Bohm interferometry analogy is most transparent for the unidirectional edge-state in the Floquet gap, a circular dichroism of the LDOS is not limited to the edge state but can be expected also in the absence of a clear gap based on the general chiral nature of the Floquet state induced by a circular drive.}

\section{Discussion}
\label{sec:discussion}
In this work, we assessed the feasibility of probing topological Floquet states in graphene with ultrafast THz-driven scanning tunneling microscopy. We developed a nonequilibrium Green's-function framework for time-dependent tunneling that directly generalizes static STM to driven and transient settings. We showed that, under circular optical driving, light-induced dynamical gaps and chiral edge states emerge in experimentally accessible observables such as rectified currents and their generalized conductances. Beyond local nonequilibrium tunneling spectroscopy, we introduced the concept of Floquet quasiparticle interference (Floquet QPI) and demonstrated how real-space scattering signals can be used to reconstruct the dispersion of coupled chiral edge states in graphene. Finally, as a ``smoking-gun'' signature of edge-state chirality, we proposed to exploit the helicity-dependent LDOS renormalization around a chiral impurity that locally breaks time-reversal symmetry. Looking ahead, Floquet QPI could be extended to access other driven quantum-geometric features---such as light-induced pseudospin textures \cite{sentef_theory_2015}---in graphene and beyond.

Realizing these ideas experimentally calls for a careful separation of genuine sample Floquet physics from competing pump-induced effects in the STM junction, including transient heating and photon-assisted tunneling (``junction dressing'') \cite{tien_multiphoton_1963,grifoni_driven_1998}. In practice, this demands quantitative control of the electromagnetic environment of the tip--sample nanogap. A central challenge is the generation of well-defined in-plane driving fields in a complex electrodynamic setting, where near-field enhancement can produce highly localized fields, strong spatial gradients, and polarization distortions. A transient THz bias can be generated reliably by illuminating the junction with weak THz fields polarized along the tip axis (out-of-plane with respect to the sample), leveraging strong antenna enhancement in the THz range \cite{walther_emission_2005,muller_phase-resolved_2020,peller_quantitative_2021}. 
In contrast, for the optical Floquet drive---and especially for circularly polarized in-plane excitation---one must control and, where necessary, suppress tip-induced near-field effects so that the intended in-plane polarization and spatial homogeneity are preserved at the sample. While the STM tip cannot be fully eliminated in an STM experiment, advanced excitation geometries (e.g.\ backside illumination at normal incidence) as well as systematic polarization control and pulse-shaping strategies provide realistic routes to reduce near-field artifacts to a minor and well-characterized role. Importantly, clean in-plane excitation is also the optimal choice to minimize parasitic junction dressing effects in the tunneling process.

\revise{This separation of sample and junction effects is supported by electromagnetic near-field simulations of the STM geometry, presented in Appendix~\ref{appx:C} For representative parameters, the simulations show that normal-incidence excitation produces no substantial near-field enhancement of either the in-plane or out-of-plane components of the optical Floquet drive at the sample. We therefore expect junction dressing from residual out-of-plane fields to be moderate, while spatial variations of the in-plane field may lead only to modest local modulations of the Floquet gap. Importantly, the topological character of the driven state is controlled by the breaking of time-reversal symmetry by the circular drive and is therefore expected to persist under moderate spatial inhomogeneity of the dressing field.}

\revise{Additional experimental considerations concern the choice of pump photon energy, field strength, and pulse duration. These parameters must be optimized to reach a sufficiently large Floquet gap while limiting junction heating and remaining compatible with the finite energy resolution currently achievable in THz-STM.}

\revise{Our simulations employ a single-particle description of the Floquet states. In principle, the tunneling formalism developed here can be extended to include Coulomb interactions, electron-phonon scattering, and other many-body effects through more realistic self-energies. Previous work suggests that, for Floquet band engineering, a primary effect of such processes is decoherence-induced linewidth broadening, which is already phenomenologically included in our simulations through the reservoir coupling. This broadening can limit the practical visibility of Floquet-induced gap openings, but it does not necessarily preclude the formation of Floquet states: recent time- and angle-resolved photoemission experiments on graphene have demonstrated Floquet band features even when decoherence times are comparable to the optical cycle \cite{merboldt_observation_2025}. Moreover, Floquet-induced, topologically trivial gaps have recently been observed in graphene under linearly polarized driving despite the comparatively limited energy resolution of photoemission relative to STM-based spectroscopy \cite{wang_observation_2026}. A more microscopic treatment of interactions and relaxation processes would therefore be valuable, but these effects are not expected to impose a fundamental limitation on the present proposal. Instead, understanding substrate embedding and electron-phonon relaxation may provide a route to engineering dissipation pathways that stabilize transient, or even quasi-steady-state, Floquet populations \cite{seetharam_controlled_2015,esin_quantized_2018,seetharam_steady_2019}.} Systematic materials optimization---including dielectric environment, disorder landscape, and heat dissipation---should therefore be viewed as an integral part of any experimental program aiming at robust, reproducible Floquet spectroscopy and interferometry in real space.

Beyond these near-term challenges, our results point to a broader roadmap in which near-field physics is not only a source of complications but also a resource. The tip--sample nanogap constitutes a tunable electromagnetic cavity whose mode structure and field confinement can, in principle, be engineered, bringing THz-STM into the context of cavity quantum materials \cite{schlawin_cavity_2022}. This perspective motivates a natural extension towards Floquet--cavity engineering and nonequilibrium fluctuation-driven regimes. Conceptually, one may envision a quantum-to-classical crossover between cavity engineering---enhanced light--matter coupling dominated by virtual photons---and Floquet engineering---free-space coupling to coherent real photons. Early theory work on this crossover in correlated lattice systems has shown that limitations of classical Floquet protocols, such as heating, decoherence, and short lifetimes, can be mitigated by cavity embedding \cite{sentef_quantum_2020,eckhardt_quantum_2022}. Remarkably, even incoherent few-photon states can reproduce Floquet-like band modifications when the effective single-particle coupling is sufficiently strong \cite{sentef_quantum_2020}, suggesting that many established Floquet ideas can be reformulated as Floquet--cavity ideas with opportunities that go beyond either the purely cavity or purely free-space Floquet setting.

If the challenges of measuring Floquet states in a THz-STM experiment can be met, the payoff could be substantial. Conceptually, THz-STM provides a direct and local probe of pump-induced LDOS renormalization with simultaneous temporal, spatial and spectral resolution, thereby complementing established momentum-space probes of driven topology. More broadly, our results position THz-STM as a versatile platform for near-field tailoring and probing of driven band topology---opening a route to spatially resolved Floquet engineering in graphene and, potentially, in a much wider class of quantum materials.

\begin{acknowledgments}
\emph{Acknowledgments.} NJ and MAS acknowledge funding from the European Union (ERC, CAVMAT, project no. 101124492). MM acknowledges funding from the European Union (ERC H2024 StG project FASTOMIC/Grant No. 101165707). Views and opinions expressed are however those of the author(s) only and do not necessarily reflect those of the European Union or the European Research Council. NJ and MAS acknowledge the hospitality of the Kavli Institute for Theoretical Physics (NSF grant PHY-2309135). MAS acknowledges funding by the Deutsche Forschungsgemeinschaft (DFG, German Research Foundation)- 531215165 (Research Unit ‘OPTIMAL’)). MS acknowledges funding from the Swiss National Science Foundation (Grant no. 10003275). This work was partially  supported by the European Research Council (ERC-2024-SyG- 101167294 ; UnMySt), the Cluster of Excellence Advanced Imaging of Matter (AIM), Grupos Consolidados y Alto Rendimiento UPV/EHU, Gobierno Vasco (IT1453-22). We acknowledge support from the Max Planck-New York City Center for nonequilibrium Quantum Phenomena. The Flatiron Institute is a division of the Simons Foundation.
Computations were performed on the HPC systems Viper and Raven at the Max Planck Computing and Data Facility. \end{acknowledgments}

\appendix

\makeatletter
% reset within each appendix section (A, B, ...)
\@addtoreset{equation}{section}
\@addtoreset{figure}{section}
% (optional) tables too:
% \@addtoreset{table}{section}
\makeatother

\renewcommand{\theequation}{\Alph{section}\arabic{equation}}
\renewcommand{\thefigure}{\Alph{section}\arabic{figure}}
% (optional) tables too:
% \renewcommand{\thetable}{\Alph{section}\arabic{table}}

% ensure clean start in Appendix A
\setcounter{equation}{0}
\setcounter{figure}{0}

\section{Real-time simulations} \label{appx:A}

\begin{figure}[htp!]
    \centering
    \includegraphics[width=0.8\linewidth]{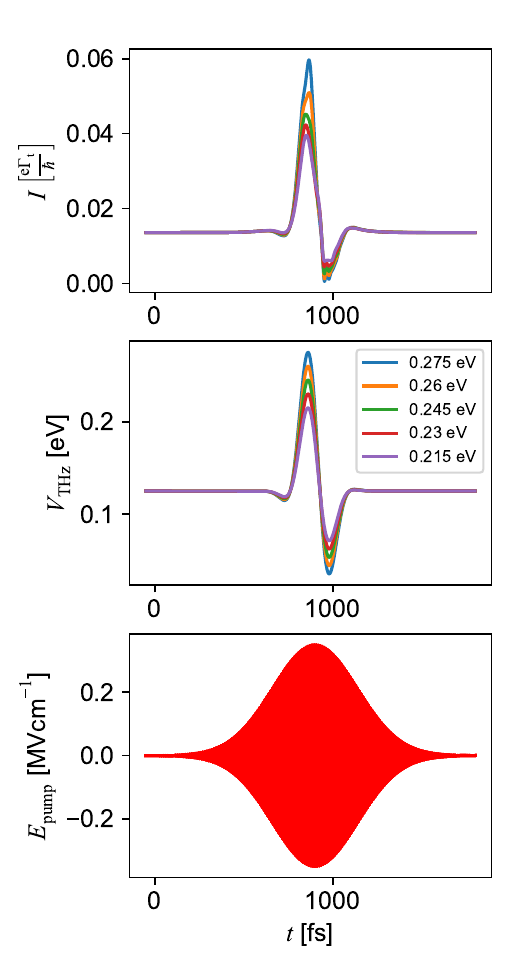}
    \caption{Time-evolution of a pump-probe protocol with the $x$ component of the pump field $E_\mathrm{pump}(t)$ bottom panel, the probe pulse $V_\mathrm{THz}(t)$ for different peak fields as indicated by the legend (center panel) and the corresponding time resolved tunneling current $I(t)$. The parameters are the ones in Sec. \ref{sec:bulkgaps}.}
    \label{fig:appx}
\end{figure}

We briefly outline the time-linear scheme used to propagate the equal-time lesser Green's function $\hat G^<(t,t)$ for a system coupled to reservoirs in the WBLA, and to evaluate the tunneling currents, as adapted from Refs.~\cite{kwok_stm_2019,tuovinen_time-linear_2023}.
The scheme relies on an efficient implementation of the lesser self-energies [Eqs.~\ref{eq:sigmasubstrate} and \ref{eq:sigmatip}], where a pole expansion is used to represent the Fermi function,
\begin{align}
    f(\varepsilon) =\frac{1}{2} - \sum_i\frac{\eta_i}{\beta} \left(\frac{1}{\varepsilon + \mathrm{i}\frac{\zeta_i}{\beta}} + \frac{1}{\varepsilon - \mathrm{i}\frac{\zeta_i}{\beta}}\right).
\end{align}
This form is reminiscent of the exact representation of $f$ as a sum over infinitely many Matsubara frequencies.
Using a Pad\'e approximation, $f$ can be represented by a finite number of poles and residues $\zeta_i$ and $\eta_i$ with $\mathrm{Re}\,\zeta_i>0$, which can be constructed numerically as described in Ref.~\onlinecite{hu_communication_2010}.
By virtue of the residue theorem, the lesser self-energies for $t>t'$ become
\begin{align}
\begin{split}
    \hat \Sigma_\alpha^<&(t,t')= \mathrm{i}  \hat \Gamma_\alpha\times\\ 
    &\left[  \frac{\delta(t-t')}{2} - \sum_n \frac{\eta_n}{\beta}e^{ -\mathrm{i}\Phi_\alpha(t,t') - \mathrm{i}\left(\mu_\alpha - \mathrm{i}\frac{\zeta_n}{\beta}\right)\frac{(t -t')}{\hbar} } \right]
\end{split}
\end{align}
for $\alpha=\mathrm{s,t}$, with $\hat \Gamma_\mathrm{s}=\Gamma_\mathrm{s}$ and $\hat \Gamma_\mathrm{t}=\Gamma_\mathrm{t}\ket{\vec r}\bra{\vec r}$.
The full time evolution can then be written as a coupled system of first-order differential equations,
\begin{widetext}
    \begin{align}
    &\mathrm{i}\hbar \frac{\partial}{\partial t} \hat G^<(t, t) = [\hat h(t),\hat G^<(t, t)] + \hat J_\mathrm{em,s}(t) +  \hat J^\dagger_\mathrm{em,s}(t) \\
    &-\mathrm{i} \hbar\frac{\partial}{\partial t'} \hat G^{A}(t, t')\left(\hat h(t') +\mathrm{i}\frac{ \Gamma_\mathrm{s}}{2}\right) =  \delta(t-t')\\
    &\hat J_{\mathrm{em},\alpha}(t) =  -\mathrm{i}\hat\Gamma_\alpha\left(\frac{\hat G^<(t,t)}{2}  - \frac{\mathrm{i}}{4} + \sum_n  \hat J_{n,\alpha}(t)  \right) \\
    &\hat J_{n,\alpha}(t) = \int\mathrm{d}t'\frac{\eta_n}{\beta} e^{ -\mathrm{i}\Phi(t,t') -\mathrm{i} \left(\mu - \mathrm{i}\frac{\zeta_n}{\beta}\right)(t-t')}\hat G^A(t',t) \\
    &\frac{\partial\hat J_{n,\alpha}(t)}{\partial t} = \hat J_{n,\alpha}(t) \left(\mathrm{i}\left(\hat h(t) -\mu_\alpha-V_\alpha(t)\right) -\frac{\Gamma_\mathrm{s}}{2} - \frac{\zeta_n}{\beta}\right) +\mathrm{i} \frac{\eta_n}{\beta}.
\end{align}
\end{widetext}
Here we neglect the backaction of the tip on the sample dynamics, as discussed in the main text; the tip embedding terms are included only to evaluate the tunneling current at each time step.

We solve the resulting system using a fourth-order Runge--Kutta algorithm. As an initial condition, $\hat G^<(t,t)$ and the embedding terms $\hat{J}_{\mathrm{emb},\alpha}$ are chosen in equilibrium with respect to the undriven Hamiltonian.
We use a time step of $\delta t = 5\times10^{-2}~\mathrm{fs}$ to $\delta t = 6\times10^{-2}~\mathrm{fs}$ and $N_\mathrm{pole}=80$ poles to represent the Fermi function.
For the bulk simulations we employ a uniform $k$ grid with $350^2$ points for $(k_x,k_y)\in\left[-0.2\frac{\pi}{a},0.2\frac{\pi}{a}\right]^2$ centered around the $K$ points, and for the zigzag ribbon simulations we use a uniform grid of 128 points for $k_x\in\left[-\frac{\pi}{a},\frac{\pi}{a}\right)$.
A representative real-time pump--probe evolution, including the pump field, probe field, and time-resolved current, is shown in Fig.~\ref{fig:appx}.

\section{Floquet calculations} \label{appx:B}
The bulk Floquet calculations are performed by direct evaluation of the Floquet retarded and lesser Green's functions, and of the corresponding spectral functions, using the Floquet Hamiltonian formulated in extended (replica) space.

For Floquet band-structure and LDOS calculations in semi-infinite graphene ribbons, we include an embedding self-energy $\mathbf{\Sigma}^R_{\mathrm{term}}$ that models embedding into an environment. Consistency requires this environment to be a driven semi-infinite graphene sheet, which leads to the self-consistent definition
\begin{align}
    \mathbf{\Sigma}^R_\mathrm{term}(\varepsilon) = \mathbf{t} \frac{1}{\varepsilon + \mathrm i \Gamma - \mathbf h - \mathbf{\Sigma}^R_\mathrm{term}(\varepsilon)}\mathbf{t}^\dagger \label{eq:termination}
\end{align}
where $\mathbf{t}$ denotes the hybridization between the terminal carbon chain and the environment. Eq.~\eqref{eq:termination} is solved iteratively in reciprocal space.

In real space, the retarded Green's function in the presence of a defect potential $\mathbf{v}$ that breaks translational invariance is
\begin{align}
    \mathbf{G}^{R}(\varepsilon) = \frac{1}{\varepsilon -\mathbf{h}  - \mathbf{v} - \mathbf{\Sigma}^R(\varepsilon)} \label{eq:defectGF}.
\end{align}
In this work, $\mathbf{v}$ represents either a hard-wall potential or additional chiral hopping terms; in both cases it is included in all Floquet replica sectors.

The hard-wall potential is implemented in the T-matrix formalism where the correction to Green's function reads 
\begin{align}
    \mathbf{G}^{R}(\varepsilon) = \left [\mathbf{G}_0^{R} + \mathbf{G}_0^{R} \mathbf{T}  \mathbf{G}_0^{R}\right](\varepsilon)
\end{align}
with the T-matrix
\begin{align}
\begin{split}
    \mathbf{T}(\varepsilon) &= \frac{1}{1 -  \mathbf{v}\mathbf{G}_0^R(\varepsilon)}\mathbf{v}\\
                            &= \ket{0}\frac{1}{1 -  \mathbf{v}_\infty \mathbf{G}_0^R(\varepsilon, x = 0) }\mathbf{v}_\infty \bra{0}.
\end{split}
\end{align}
The last line is valid for defects within a single unit cell. 
For an appropriate choice of the ribbon unit cell, the hard-wall potential can be defined as $\mathbf v = \mathbf{v}_\infty \ket{0}\bra{0}$, where $\ket{0}$ projects to the position $x = 0$ whereas $\mathbf{v}_\infty$ compromises the values of the defect potential along the $y$ direction as well as the Floquet replica structure. 
Note that due to periodic boundary conditions, this form of potential implies a restriction of the geometry from both sides along $x$. 
The correction to the Green's function can then be evaluated as
\begin{align}
        \delta&\mathbf{G}^R(\varepsilon, x) = \mathbf{G}^R_0(\varepsilon, x)\frac{1}{1 -  \mathbf{v}_\infty  \mathbf{G}_0^R(\varepsilon, x =0) }\mathbf{v}_\infty\mathbf{G}^R_0(\varepsilon,- x).
\end{align} 
and requires only the the computation of Green's function in $k$ space $\mathbf{G}^R_0(\varepsilon, k_x)$ and its Fourier transform $\mathbf{G}^R_0(\varepsilon, x )$ instead of the full real-space computation. 

The Floquet structure of $\mathbf{T}$ implies that scattering processes in all Floquet sectors contribute to the time averaged LDOS in the zeroth Floquet subspace.
These additional contributions distinguish Floquet QPI from conventional QPI and can be intuitively understood as impurity scattering events happening in between the photon absorption and emission processes of quasiparticles.

The calculations involving chiral defects are conducted with a computationally costly but feasible direct real-space implementation of the Green's function according to Eq.~\ref{eq:defectGF}, where the Floquet Hamiltonian must be evaluated based on the real-space tight-binding Hamiltonian for graphene,
\begin{align}
    \hat{h}(t) = -\gamma_0\sum_{\langle ij\rangle}  e^{\mathrm{i} (\vec R_j - \vec R_i)\cdot  \frac{\mathrm e \vec A(t)}{\hbar } }\ket{i}\bra{j}.
\end{align}
The sum runs over all pairs of nearest neighbor sites $\langle i j\rangle$ on the honeycomb lattice with the respective connection vectors $\vec R_j - \vec R_i$ entering in the Peierls phase.

We checked convergence of all Floquet computations and found replica cut-offs $\leq 10$ to be sufficient in all reported cases.

\section{Tip-induced near fields} \label{appx:C}
\begin{figure*}
\centering
\includegraphics[width=\linewidth]{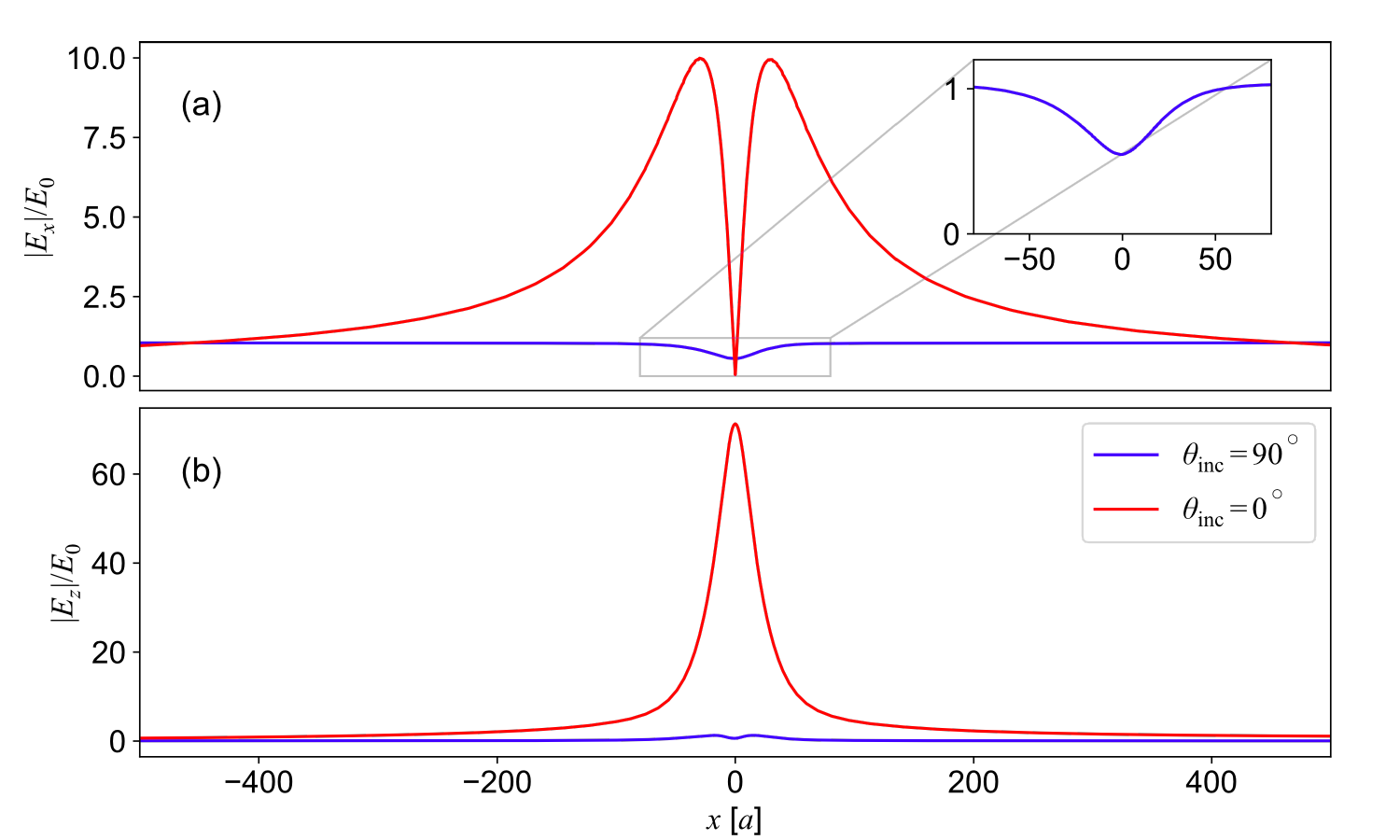}
\caption{\revise{Lateral near-field distributions of the (a) in-plane and (b) out-of-plane electric-field components in freestanding graphene placed 1nm from the tip apex. The junction is excited with linearly polarized light at a wavelength of $3\mu\mathrm{m}$ under normal incidence (blue curves). For comparison, the red curves show the field enhancement for side illumination with linear polarization parallel to the tip axis. The $x$ axis is given in units of the graphene lattice constant, $a=2.46~\mathrm{\AA}$. The electric-field components are normalized to the incident field strength.}}
\label{fig:nearfield}
\end{figure*}

\revise{
Fig.\ref{fig:nearfield} shows the lateral near-field distribution inside a freestanding graphene sheet in the presence of an STM tip under plane-wave illumination at a wavelength of $\lambda=3\mu\mathrm{m}$. The incident field is linearly polarized at incidence angles of $\theta_\mathrm{inc}=0^\circ$ (side illumination with tip-parallel polarization) and $\theta_\mathrm{inc}=90^\circ$ (backside illumination at normal incidence with in-plane polarization). The field is computed using a finite-element simulation in COMSOL Multiphysics. The tip is made of tungsten, with an apex radius of 30nm, and the tip–sample distance is set to 1nm.
The freestanding graphene is modeled using a transition boundary condition with an effective thickness of 0.34nm and an electrical conductivity of $1.79\times10^5\mathrm{S,m^{-1}}$ at the surface of a thick vacuum layer. The electric field is evaluated 0.1~nm above and below the graphene sheet and subsequently averaged to estimate the electric field inside the graphene layer. For the fully symmetric geometry at normal incidence, the field distribution is symmetrized with respect to the tip center to remove small residual numerical asymmetries.
In the case of normal incidence ($\theta_\mathrm{inc}=90^\circ$), the in-plane field $E_x$ is reduced by approximately a factor of two relative to the incident field $E_0$ and varies on a characteristic length scale of roughly 50–100 lattice constants for the tip radius considered here. This length scale is expected to be sufficiently large to support well-defined local Floquet states. In addition, the near field exhibits an out-of-plane component $E_z$ with a magnitude comparable to that of the incident field in the vicinity of the tip. This component may give rise to additional effects, such as photon-assisted tunneling. The spatial extent and degree of inhomogeneity of the tip-induced field redistribution depend on the tip radius, shaft opening angle, and precise illumination angle. This dependence may allow for further optimization of the experimental conditions using advanced tip-fabrication methods, such as tip shaping by focused-ion-beam milling.
Notably, these near-field effects are significantly weaker than for $\theta_\mathrm{inc}=0^\circ$, for which the incident electric field is aligned parallel to the tip axis, resulting in strong field enhancement and a highly inhomogeneous field distribution.
A more comprehensive investigation of the influence of the tip-induced near field on Floquet states is left for future work. Nevertheless, the present simulations suggest that, under normal incidence, the field redistribution introduced by the STM tip may quantitatively modify the Floquet response without altering the central conclusion that Floquet states should remain experimentally accessible in THz-STM measurements.
}

% Bibliography
% \bibliographystyle{apsrev4-2}
\bibliography{references.bib}

\end{document}